\documentclass[journal]{vgtc}                % final (journal style)
\usepackage{graphicx}
\usepackage{times}
\usepackage{todonotes}
\usepackage{enumerate}

\usepackage[bookmarks,backref=true,linkcolor=black]{hyperref} %,colorlinks
\hypersetup{
  pdfauthor = {},
  pdftitle = {},
  pdfsubject = {},
  pdfkeywords = {},
  colorlinks=true,
  linkcolor= black,
  citecolor= black,
  pageanchor=true,
  urlcolor = black,
  plainpages = false,
  linktocpage
}
\DeclareGraphicsExtensions{.pdf,.jpg,.jpeg} % for pdflatex we expect .pdf, .png, or .jpg files
\usepackage{cite}                      % needed to automatically sort the references
\usepackage{tabu}                      % only used for the table example
\usepackage{booktabs}                  % only used for the table example
\usepackage{graphicx}
\usepackage{subcaption}
\usepackage{fancyvrb}
\usepackage{listings}		% support for listings
\usepackage{array}
\usepackage{booktabs}
\usepackage{hyphenat}
\usepackage{theorem}
\usepackage{amssymb}
\usepackage{amsmath} 
\usepackage{ctable}
\usepackage{url}

\usepackage{multirow}
\usepackage{sidecap}
\usepackage{placeins}
\usepackage{color,colortbl}					% colors used in listings
\usepackage{afterpage}

\usepackage{algorithm}
\usepackage{algpseudocode}

\definecolor{RED}{rgb}{1,0,0}

\def\mathbi#1{\textbf{\em #1}}
\newcommand{\eps}{\mbox{$\epsilon$}}

\def\x{\mathbi{x}}

\let\vec=\mathbi%
\let\mat=\mathbf%
\let\set= \mathcal%

\DeclareMathOperator*{\argmin}{arg\,min}

\onlineid{0}

\vgtccategory{Research}
\vgtcpapertype{please specify}

\title{Visualization of High-dimensional Scalar Functions \\ using Principal Parameterizations}

\author{Rafael Ballester-Ripoll, \textit{Member, IEEE} and Renato Pajarola, \textit{Senior Member, IEEE}}
\authorfooter{
\item
 R. Ballester-Ripoll and R. Pajarola are with the University of Zurich, Department of Informatics. E-mails: \{rballester,pajarola\}@ifi.uzh.ch.
}

\shortauthortitle{Ballester-Ripoll \MakeLowercase{and} Pajarola: Visualization of High-dimensional Scalar Functions Using Principal Parameterizations}
\abstract{Insightful visualization of multidimensional scalar fields, in particular parameter spaces, is key to many fields in computational science and engineering. We propose a principal component-based approach to visualize such fields that accurately reflects their sensitivity to input parameters. The method performs dimensionality reduction on the vast $L^2$ Hilbert space formed by all possible partial functions (i.e., those defined by fixing one or more input parameters to specific values), which are projected to low-dimensional parameterized manifolds such as 3D curves, surfaces, and ensembles thereof. Our mapping provides a direct geometrical and visual interpretation in terms of Sobol's celebrated method for variance-based sensitivity analysis. We furthermore contribute a practical realization of the proposed method by means of tensor decomposition, which enables accurate yet interactive integration and multilinear principal component analysis of high-dimensional models.
} % end of abstract


\teaser{
  \centering
  \begin{subfigure}[b]{0.295\textwidth}
    \includegraphics[width=1\linewidth]{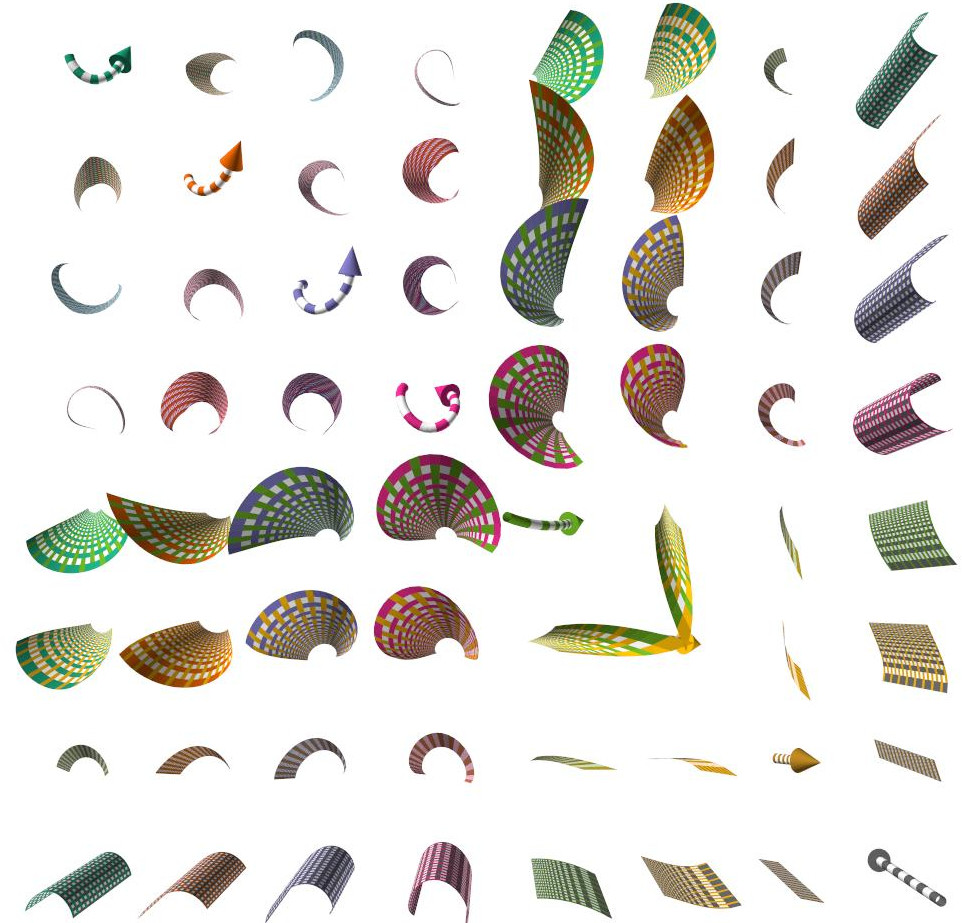}
  \end{subfigure} \hspace{0.2cm}
  \begin{subfigure}[b]{0.15\textwidth}
    \includegraphics[width=1\linewidth]{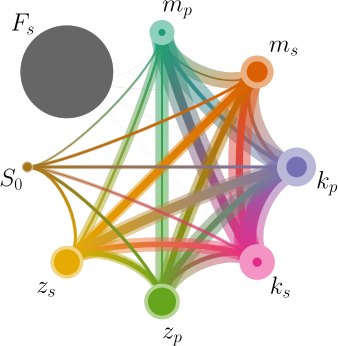}
  \end{subfigure} \hfil
  \begin{subfigure}[b]{0.02\textwidth}
    \includegraphics[width=0.0245\linewidth]{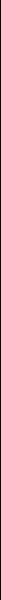}
  \end{subfigure} \hfil
  \begin{subfigure}[b]{0.295\textwidth}
    \includegraphics[width=1\linewidth]{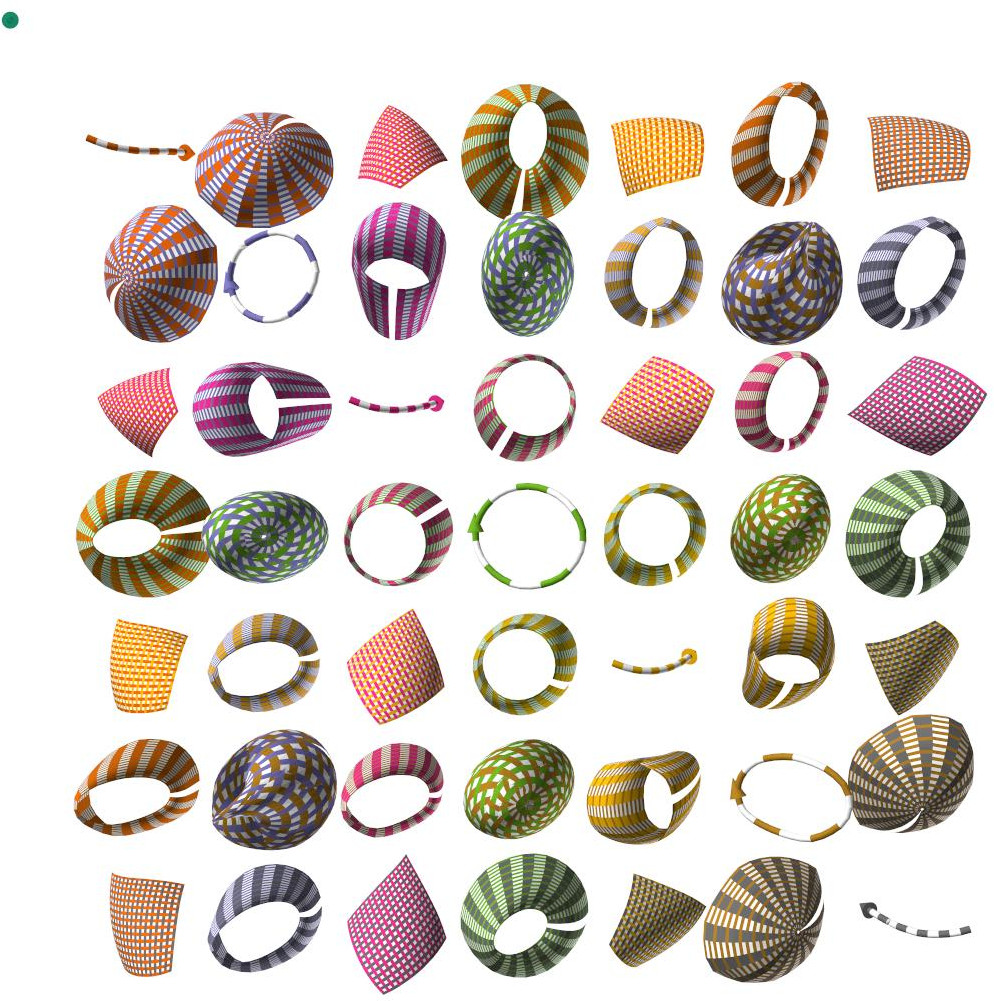}
  \end{subfigure} \hspace{0.2cm}
  \begin{subfigure}[b]{0.15\textwidth}
    \includegraphics[width=1\linewidth]{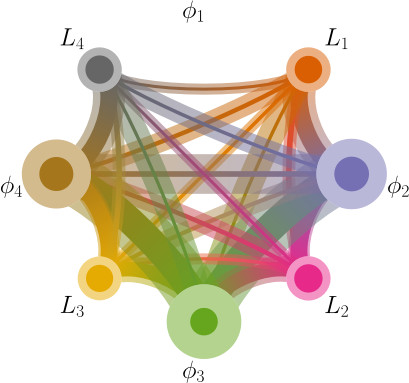}
  \end{subfigure}
  \caption{Plot matrices showing all 1st- and 2nd-order principal parameterizations for the 8-dimensional \emph{Robot Arm} and \emph{Damped Oscillator} physical models, respectively. The proposed mapping reveals non-essential variables, periodicity, inter-variable effects, etc. The small radial plot in each case summarizes the corresponding global sensitivity indices as given by Sobol's method.}
	\label{fig:teaser}
}

\begin{document}


\firstsection{Introduction}
%------------------------------------------------------------------------------------------------------------------------------

\maketitle

Dimensionality reduction is a crucial data processing step to interactively visualize and explore large complex data sets in many visual data analysis methods and systems. Past efforts in the field, including those based on linear projections, are largely tailored for scattered data visualization. Such data sets comprise a finite amount of available data samples that are expected to follow an often unknown pattern; hopefully, an interesting manifold that explains well the sampled data points' distribution within their high-dimensional domain. Much of the visualization challenge therein consists of learning and revealing this low-dimensional underlying structure.

However, many important problems are actually \emph{dense} in nature. For example, parameter spaces in engineering and life sciences or econometric models (including black-box systems, simulations, and metamodels) are often high-dimensional and may accept a large or even infinite number of valid parameter combinations. General scattered data approaches are less effective for such situations. Parameter space analysis and multidimensional scalar function visualization have thus become a research field of their own, for which specialized techniques have been introduced. How to perform dimensionality reduction in this scenario, as is customary for the scattered case, is still a major challenge. The usual \emph{linear} vs. \emph{non-linear} distinction~\cite{SBIM:12} still applies, but this dense setting brings about some new specific visualization challenges and use cases~\cite{SHBPM:14}.

In this paper we pursue the goal of meaningfully depicting a model's dependence with respect to its input variables as well as joint interactions of arbitrary groups of variables. Critically, we also want to understand how the model's behavior \emph{evolves} as these variables (and groups thereof) take different specific values. The dimensionality reduction we propose is inspired by the popular \emph{Sobol method} for sensitivity analysis~\cite{Sobol:90}, which partitions a model into orthogonal projected subfunctions that depend on different subsets of variables. It is an example of \emph{analysis of variance} (ANOVA) method that has long attracted great interest in uncertainty quantification and reliability engineering.

Consider a domain $\Omega = \Omega_1 \times \dots \times \Omega_N \subset \mathbb{R}^N$ over which $N$ variables $x_1, \dots, x_N$ move and a multidimensional function $f: \Omega \to \mathbb{R}$. Given some variables of interest $x_{i_1}, \dots, x_{i_K}$, we introduce the \emph{principal parameterization} with respect to these variables as a mapping $\pi: \Omega_{i_1} \times \dots \times \Omega_{i_K} \to \mathbb{R}^D$ that is \emph{as similar as possible} to the original $f$. We will detail the precise desired notion of similarity in the central sections of this paper. $D$ is the target reduced dimensionality; w.l.o.g. here we will always use $D = 3$ for effective visualization, namely an embedding into a 3D system of coordinates. Hence, for $K=1$ (one variable of interest) we produce parameterized curves in $\mathbb{R}^3$, for $K=2$ surfaces (or equivalently, ensembles of curves), for $K=3$ volumes (or ensembles of surfaces), etc.

As opposed to strategies that focus mainly on certain critical or topologically interesting points (for example, local extrema for Morse-Smale complexes~\cite{GBPW:10}), our approach is a dimensionality reduction that takes the \emph{full} model $f$ and its exponentially large domain $\Omega$ into consideration, not unlike e.g. the active subspace method~\cite{Constantine:15}. This kind of catch-all approaches are good at conveying context within a model and are thus attractive for the so-called \emph{global-to-local} user navigation paradigm~\cite{SHBPM:14}. However, they have been exploited to a limited degree only due to the \emph{curse of dimensionality}. For instance, one often needs to compute multidimensional integrals over the entire $\Omega$. This is an intensively researched and computationally expensive task, and is all the more challenging with the computing time limitations that usually arise in a visualization context. As a second contribution we make the proposed system computationally feasible and real-time responsive by means of a convenient compact representation, namely a \emph{tensor decomposition}~\cite{Oseledets:11}. That way, one can efficiently manipulate dense high-dimensional scalar functions and reconstruct (decompress) regions of interest or derived properties interactively. In principle, the proposed transformation could also be computed using another back-end numerical framework, for instance (quasi-)Monte Carlo or sparse grid-based approximate integration.

%------------------------------------------------------------------------------------------------------------------------------
\subsection*{Notation}

Given a function of interest $f:\Omega \subset \mathbb{R}^N \to \mathbb{R}$, we will refer to its \emph{partial functions} (or just \emph{partials}) as the functions that arise by fixing some $K \ge 1$ of $f$'s variables. For example, $f^{x_1}$ is an $(N-1)$-variate function defined as $f^{x_1=\alpha}(x_2, \dots, x_N) := f(x_1=\alpha, \dots) = f(x_1=\alpha, x_2, \dots, x_N)$. Whenever we wish to identify one of these subfunctions we will simply write $f^{x_n}$. For $K \ge 2$ we have higher-order partials that fix two or more variables, for example $f^{x_n, x_m}$.

Tensors in this context are multidimensional data arrays, including vectors, matrices, volumes and so forth. Vectors and matrices are denoted using bold italic lowercase ($\vec{u}$) and roman uppercase ($\mat{U}$) respectively, whereas general tensors use calligraphic letters ($\set{T}$).

%------------------------------------------------------------------------------------------------------------------------------
\section{Related Work} \label{sec:related_work}
%------------------------------------------------------------------------------------------------------------------------------

%------------------------------------------------------------------------------------------------------------------------------
\subsection{Parameter Space Visualization}

Several visualization frameworks lend themselves well to parameter space analysis. These include continuous parallel coordinates~\cite{HW:13}, dimensional stacking~\cite{WLT:94}, the HyperSlice~\cite{WL:93} and Sliceplorer~\cite{TSM:17} tools, etc. Others are highly domain-specific, such as Tuner~\cite{TSMHWV:11} for brain segmentation or~\cite{BM:10} for physically-based simulations in graphics. In terms of the conceptual framework defined by the comprehensive survey by Sedlmair et al.~\cite{SHBPM:14}, our proposed approach is geared towards global-to-local exploration and places a special emphasis on the \emph{sensitivity analysis} task. We refer the reader to~\cite{SHBPM:14} for a more inclusive literature overview and limit this section to our particular scope and use cases.

In sensitivity-oriented visualization there is a parameter space $f(x_1, \dots, x_N)$ whose variables are either freely tunable by the user (usually in a controlled experimental or simulated environment) or naturally governed by a probability density function (PDF). In the latter case, the complexity that is due to the model function $f$ adds to that of its underlying PDF, which may or may not be known in closed form. If one wishes to place a strong emphasis on the PDF, one may take a set of representative samples distributed accordingly and then simply apply their favorite scattered data visualization technique~\cite{SBIM:12,LMWBP:17}. Conversely, if only a set of scattered samples is known from an otherwise dense parameter space, a \emph{surrogate model} may be fitted in order to estimate the true model during interactive visualization. This strategy provides more contextual information than a bare-bones collection of scattered points because a surrogate can be evaluated cheaply at previously uncharted locations within the domain. This enables, among others, derivative-based feature visualization in points' neighborhoods (gradients, extremal/saddle structure and other local properties) using for example flow-based scatterplots~\cite{CCM:10} or multiscale lensing~\cite{SMSL:17}.

Here we focus on the case where the PDF is of limited interest or even uniform (and especially, when parameters may be set at will). Note that this is a common scenario in sensitivity analysis. In other words, we are concerned with understanding and visualizing the complexity ascribed to the high-dimensional model $f$ itself, rather than to its parameters' distribution. A popular approach is to track and visualize $f$'s topological properties; watershedding segmentation, Morse-Smale complexes~\cite{GBPW:10} and topological spines~\cite{CLB:11} belong to this paradigm. Such methods are very sensitive to high-frequencies and irregularities in the model, and they often resort to a smoothing hyperparameter to filter out noise and reveal topological features at different scales. 
The active subspace method~\cite{Constantine:15}, which is very similar to the structure tensor idea for images and volumes~\cite{Knutsson:89}, is perhaps one of the closest to the present work: it is also based on extracting principal directions of variance within an $L^2$ space and inner product. However, while active subspaces arise from uncentered covariances between the model's gradient components $\frac{\partial f}{\partial x_1}, \dots, \frac{\partial f}{\partial x_N}$ across the domain $\Omega$, our method uses the covariance between all partial functions, be it of single variables or variable tuples. In particular, we are not limited solely to global structure. Rather, we can look at variations that occur as one or more input parameters evolve. This is a crucial feature that facilitates effective global-to-local navigation as motivated earlier.

%------------------------------------------------------------------------------------------------------------------------------
\subsection{Sobol Method for Sensitivity Analysis}  \label{sec:sensitivity_analysis}

Several decades after its inception, Sobol's method~\cite{Sobol:90} remains one of the most prominent for sensitivity analysis of high-dimensional scalar functions~\cite{SRACCGST:08, FRM:13}. Its main insight is to realize that every variable's influence can be decomposed into two orthogonal components:

\begin{itemize}
	\item The \emph{additive} (or \emph{first-order}) term $S_n$ of a variable $x_n$ measures how strongly the model's average is affected when $x_n$ moves. A purely additive variable $x_n$ means that we can separate it from $f$ and write
	\begin{equation}
	f = g(x_n) + h(x_1, \dots, x_{n-1}, x_{n+1}, \dots, x_N).
	\end{equation} 
	
	\item The \emph{high-order} term of a variable $x_n$ measures its impact on the model that is not attributable to changes of its average. Hence, a purely high-order variable $x_n$ means that the function's expected value $\mathbb{E}[f^{x_n}]$ is not affected as $x_n$ moves.
\end{itemize}

Often, these two components show up together. Their aggregation is the so-called \emph{total effect}, denoted as $S^T_n$. The first-order index $S_n$ gives a precise measure of its variable's isolated effect, but disregards joint interactions altogether. On the other hand, the total effect $S^T_n$ accounts for such interactions, although it does not identify what orders of interactions are prevalent and how partner variables are interacting. Note that the Sobol components are defined for \emph{tuples} of variables as well.

Sobol's method is excellent at robustly capturing joint relationships among interacting groups of variables. For example, while the presence of a strongly additive variable may destroy the local extrema of an otherwise topologically rich function, it will not alter the Sobol relative importances between the remaining variables. However, a drawback of the method is that for each variable (or tuple thereof), its effect over the entire domain is summarized into a scalar quantity only. Thus, it fails at reflecting changes at different \emph{specific} values of these variables. For instance, a variable may play an overall important role, but only when it takes extreme values, or it may be additive in an interval and high-order in another. The need to convey this important, more granular information calls for a novel visualization-oriented methodology that interplays well with the principles of Sobol's ANOVA framework. This is the main motivation behind our method.

%------------------------------------------------------------------------------------------------------------------------------
\subsection{Tensor Metamodeling and Visualization}

The dimensionality reduction technique we propose is based on PCA projection in vector spaces of very high dimensionality. To cope with this computational challenge we will use a framework known as \emph{tensor decomposition} that we briefly review here.

Tensor decompositions approximate arbitrary data tensors (multidimensional arrays) as expansions of simpler, separable tensors that can cope well with the curse of dimensionality~\cite{KB:09}. In the context of surrogate modeling, tensors are often defined as discretizations of parameter spaces over regular grids, whereby each tensor entry corresponds to one particular combination of parameters. For instance, given a simulation depending on $N = 8$ parameters, we may discretize each axis into 64 possible values to yield a tensor with $64^{8} \approx 3 \cdot 10^{14}$ elements. It is often possible to handle such massive tensors succinctly using a suitable decomposition, so that they never need to be managed in a raw explicit form. In this paper we use the \emph{tensor train} (TT) model~\cite{Oseledets:11}, which in recent years has been used increasingly for surrogate modeling and visualization~\cite{VDSL:14, BPP:16, GJ:18} as well as for sensitivity analysis~\cite{BEM:16, BPP:17, BPP:18}. Tensor model fitting is an active research field, and multiple options exist nowadays for either a given set of training samples or when new data points can be sampled on demand. Here we follow a precomputed metamodeling paradigm: the surrogate is built offline, taking as many samples as needed to ensure a sufficiently low estimated generalization error. No further samples are acquired during visualization, and in particular no \emph{steering} is considered.

The techniques we present take advantage of the unique strengths of tensor decompositions and, in particular, the TT model. Classical regressors such as Gaussian processes (kriging) or radial basis functions are popular for surrogate modeling in certain cases, e.g. when the available ground-truth samples are fixed and very limited in number. Nonetheless, they are less adequate for the kind of multidimensional integration and multilinear PCA projection required by the proposed visualization. The general idea of using compressed tensor coefficients as features for model (post-)processing and analysis is not new~\cite{WXCLWY:08, IVWJ:17}, and PCA is a long-established framework that has been used in the past for low-dimensional parameterization (prominently, trajectory curves over time~\cite{GD:03,CBR:16}). However, the proposed sensitivity-aware dimensionality reduction for dense, high-parametric models is new. We cover it in detail over the next sections.

%------------------------------------------------------------------------------------------------------------------------------
\section{Proposed Dimensionality Reduction} \label{sec:dimensionality_reduction}
%------------------------------------------------------------------------------------------------------------------------------

Consider an $N$-dimensional parameter space represented as a function $f:\Omega \to \mathbb{R}$. In the multivalued case $f:\Omega \to \mathbb{R}^M$ one can handle each output $1, \dots, M$ separately or, if a joint analysis for all outputs is desired, reduce the problem to the single-valued version by stacking all outputs to form an extra dimension: $f:\Omega \times \{1, \dots, M\} \to \mathbb{R}$. For simplicity, let us also assume that all inputs are continuous and scaled to $[0, 1]$ (alternative cases work analogously).

%------------------------------------------------------------------------------------------------------------------------------
\subsection{Single Variable Case}

For the sake of clarity, let us start with $K=1$, i.e. there is only one variable of interest $x_n$. Our goal is to understand its effect on the high-dimensional function $f$ or, in other words, the relationship between the $(N-1)$-dimensional partial functions $f^{x_n} = f(\dots, x_n=\alpha, \dots)$ as $\alpha$, thus $x_n$ moves between $0$ and $1$. Of course, each partial may have a structure (almost) as complex as the original $f$, so their joint behavior is just as potentially intricate and challenging.

We propose to consider the $L^2$ space $\set{F}$ of all functions that map $[0, 1]^{N-1}$ to $\mathbb{R}$. Clearly, $f^{x_n} \in \set{F}$ for every $0 \le x_n \le 1$. Let us summarize this collection of partial functions as a parameterized curve, i.e. map each to a point in $\mathbb{R}^3$. We start by averaging each partial over its free variables to get a single scalar, i.e. computing the global average of $f^{x_n}$ for any fixed $x_n$. To this end, we consider the \textit{projection function} $f_n$:

\begin{equation}
\begin{split}
f_n(x_1, \dots, x_N) := \mathbb{E}[f^{x_n}] = \\
= \int_{[0, 1]^{N-1}} f(x_1, \dots, x_N) \, dx_1 \dots dx_{n-1} dx_{n+1} \dots dx_N.
\end{split}
\label{eq:projection}
\end{equation}

The projection $f_n$ is constant along all variables but $x_n$. It captures the aggregated additive behavior as per variable $x_n$ and hence determines the first-order Sobol index (Sec.~\ref{sec:sensitivity_analysis}), which is defined as $S_n := \mathrm{Var}[f_n] / \mathrm{Var}[f]$. Such an axis-aligned projection has already some visualization power and has been used in prior literature, but obviously still gives limited information on $f^{x_n}$'s inner workings as $x_n$ varies. The missing information is contained in the \textit{corrected function} $f_{-n} := f - f_n$, which is the main idea driving Sobol's ANOVA method. It removes the additive component that is due to $x_n$ and gives rise to the total index: $S^T_n := \mathrm{Var}[f_{-n}] / \mathrm{Var}[f] + S_n$.

We now extend this idea to our setting, which is concerned with partial functions $f^{x_n}$, and split each of those partials similarly as

\begin{equation*}
f^{x_n} = f_n^{x_n} + f_{-n}^{x_n}.
\end{equation*}

For any $x_n$ these two components are furthermore orthogonal with one another w.r.t. the $L^2$ inner product:

\begin{equation*}
\label{eq:orthogonal_subspaces}
\int f_n^{x_n} f_{-n}^{x_n} = \mathbb{E}[\mathbb{E}[f^{x_n}] \cdot (f^{x_n} - \mathbb{E}[f^{x_n}])] = 0 \mbox{ for all } 0 \le x_n \le 1.
\end{equation*}

This motivates us to represent the original space of partial functions in terms of two orthogonal subspaces and brings us to the core part of the proposed mapping. In order to facilitate visualization we will decompose each $f^{x_n}$ in terms of a coordinate triplet:

\begin{itemize}
	\item One coefficient $\pi_x(x_n)$ that encodes the projection of each $f^{x_n}$ onto the subspace spanned by $f_n$'s partials along $x_n$, i.e. $\{f_n^{x_n} \,|\, 0 \le x_n \le 1\}$. As argued earlier, each of those partials is a constant function. Therefore, they are all multiple of one another, and thus they form a subspace of dimension 1 within $\set{F}$. Hence, one coordinate is enough to encode the projection exactly, and it is precisely the expected value $\mathbb{E}[f^{x_n}]$ as per Eq.~\ref{eq:projection}.
	\item Two coefficients $\pi_y(x_n)$ and $\pi_z(x_n)$ that encode the projection of each $f^{x_n}$ onto the subspace spanned by $f_{-n}$'s partials, $\{f_{-n}^{x_n} \,|\, 0 \le x_n \le 1\}$. Since this subspace's dimension is in general infinite, we resort to a truncated basis expansion. We choose an optimal basis in the $L^2$ sense, namely the two leading eigenfunctions of the Karhunen-Lo\`{e}ve expansion (KLE) for the pairwise covariance function $\mathrm{Cov}(\alpha, \beta) := \mathbb{E}[f_{-n}^{x_n = \alpha} \cdot f_{-n}^{x_n = \beta}]$ for all $0 \le \alpha, \beta \le 1$.
\end{itemize}

In summary, each individual $f^{x_n}$ is reduced to a point in 3D $(\pi_x(x_n), \pi_y(x_n), \pi_z(x_n))$ and thus the set of all $x_n \in [0, 1]$ is mapped to a parameterized curve in $\mathbb{R}^3$. This is a subspace-constrained KLE: we force a specific vector to appear in the basis, and want to find others that best summarize the remaining subspace that is orthogonal to that vector. The fixed vector gathers \emph{absolute} information, since coordinate $\pi_x$ equals the partial function's mean. On the other hand, the two other vectors to be sought encode \emph{relative} information, as absolute positions $(\pi_y, \pi_z)$ on the $yz$-plane are not directly interpretable, but the distances between points are. Although the fixed vector is not generally one of the KLE's leading eigenfunctions, it is still a reasonable basis choice in $L^2$ terms and it often captures a significant amount of the model's variance.

%------------------------------------------------------------------------------------------------------------------------------
\subsection{Multivariate Case}

We have just mapped a single variable's corresponding collection of partial functions onto a 3D curve ($K = 1$-dimensional manifold). The higher-dimensional case ($K \ge 2$) follows naturally from that. The main difference is that we have now collections that are indexed by two or more variables, and we handle higher-order partial functions, e.g. $f^{x_n, x_m}$ for $K = 2$, that arise from fixing several variables and are $(N-K)$-dimensional. As a result we no longer obtain parameterized curves but higher-order manifolds (surfaces, volumes, etc.) that are parameterized by triplets of multidimensional functions $\pi_x, \pi_y, \pi_z: [0, 1]^K \to \mathbb{R}$.

%------------------------------------------------------------------------------------------------------------------------------
\section{Practical Algorithm}
%------------------------------------------------------------------------------------------------------------------------------

%------------------------------------------------------------------------------------------------------------------------------
\subsection{Algorithm Outline} \label{sec:outline}

For the ease of exposition, we assume that the $K$ variables of interest are the first $1, \dots, K$. The dimensionality reduction that we motivated in Sec.~\ref{sec:dimensionality_reduction} boils down to a three-step processing pipeline:

\begin{enumerate}[{Stage }A:]
\item The \emph{within-mean} of each partial function of $f$ is computed (to be used as $x$-coordinate during visualization) and subtracted from the original. This way we derive a new function $f_{-1 \dots K}$ whose partials are zero-centered.
\item The \emph{cross-mean} of $f_{-1 \dots K}$ (i.e., its average over target variables $1, \dots, K$) is subtracted from each of its elements to yield a new collection $f_{\overline{-1 \dots K}}$. By doing so we are shifting the collection's origin of coordinates to its barycenter as is often done in PCA to achieve a more meaningful projection.
\item We compute the two leading eigenfunctions of the $[0, 1]^K \times [0, 1]^K \to \mathbb{R}$ covariance function that maps every possible pair of elements of $f_{\overline{-1 \dots K}}$ to their inner product. For each partial, its coefficients in terms of this basis define its embedding on the $yz$-plane. 
\end{enumerate}

\begin{figure}[h]
\centering
\begin{subfigure}[c]{0.31\columnwidth}
\includegraphics[width=1\columnwidth]{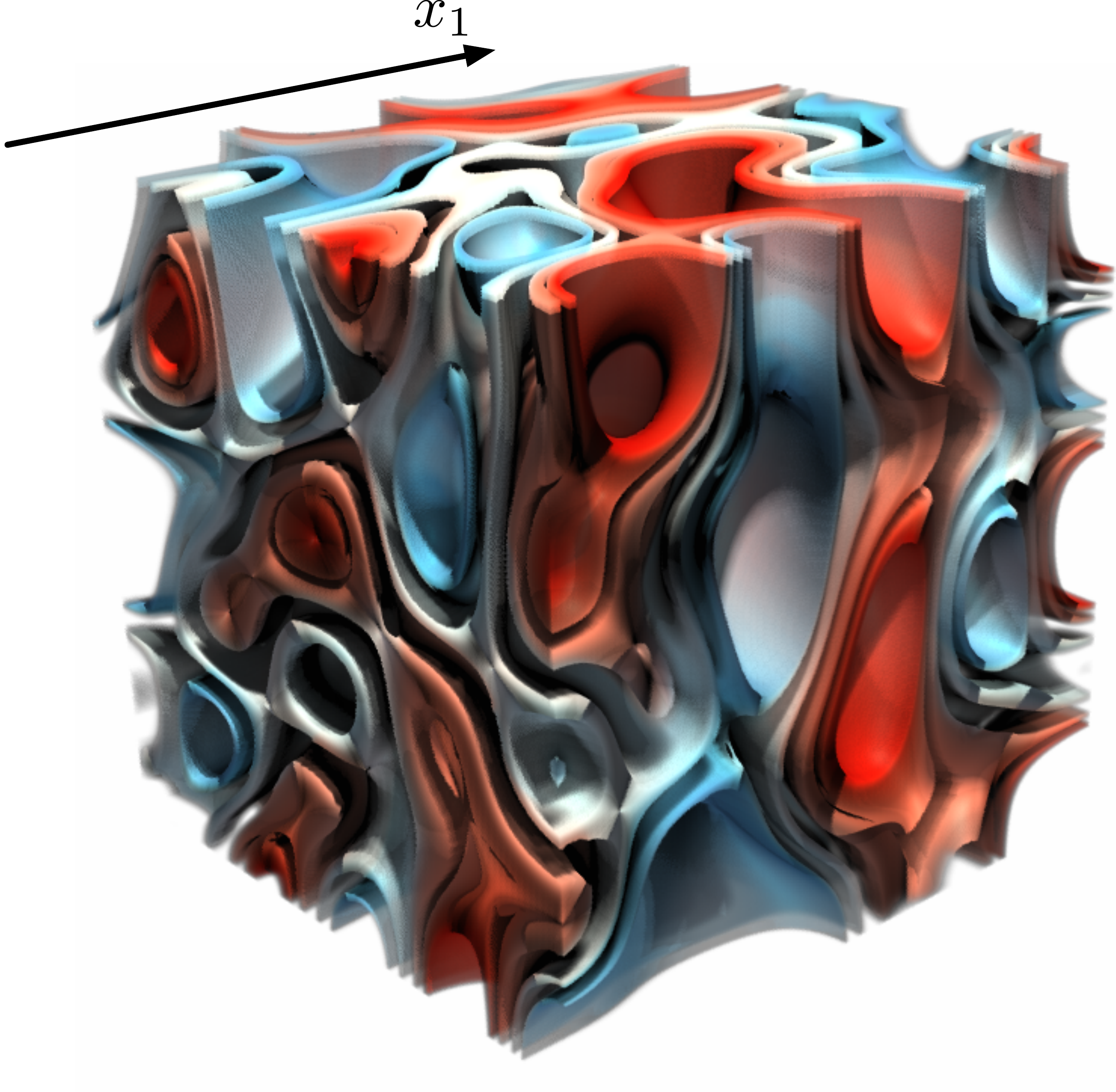}
\caption{Original $f$}
\end{subfigure} \hfil
\begin{subfigure}[c]{0.31\columnwidth}
\includegraphics[width=1\columnwidth]{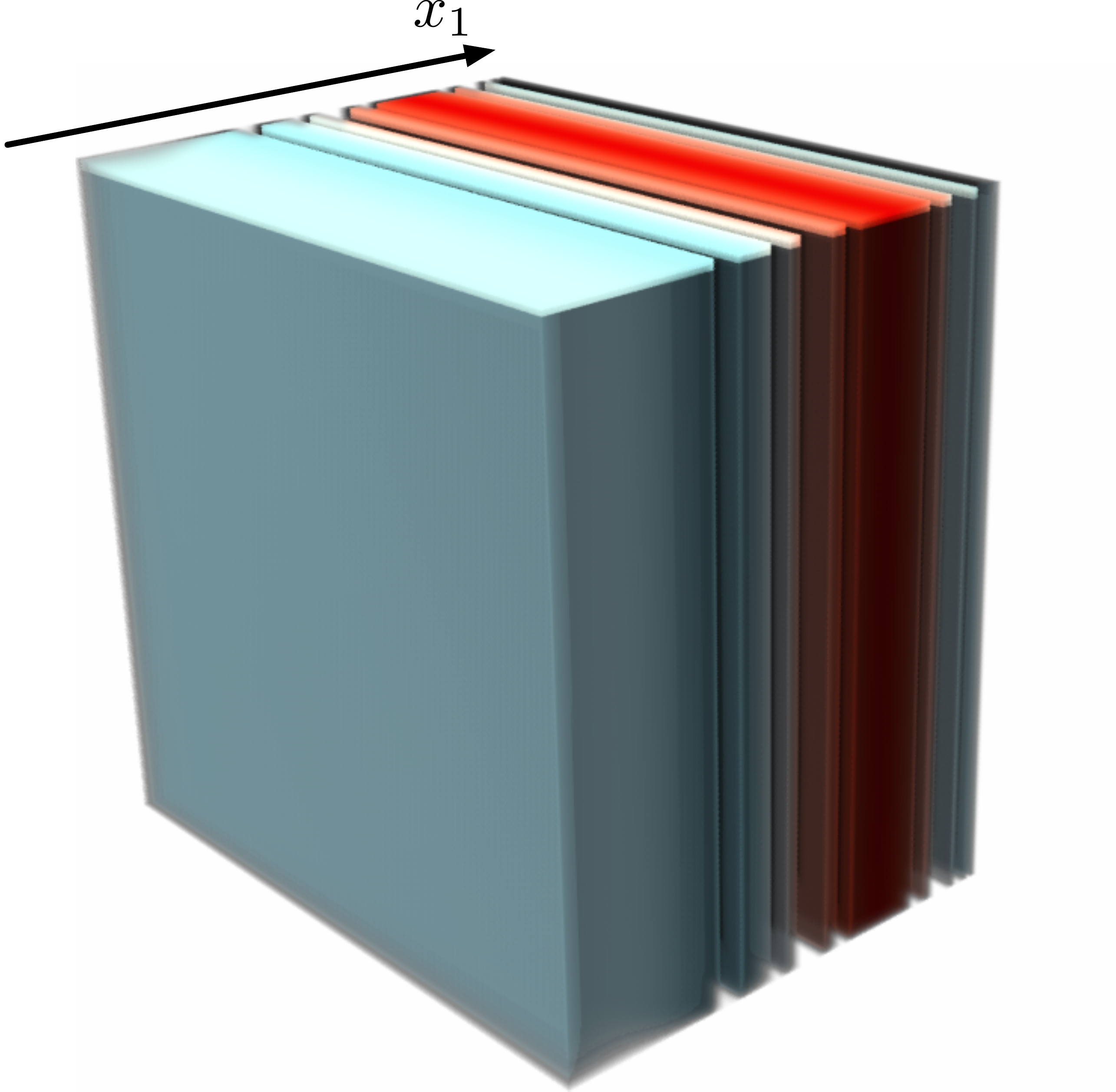}
\caption{Within-mean $f_1$}
\end{subfigure} \hfil
\begin{subfigure}[c]{0.31\columnwidth}
\includegraphics[width=1\columnwidth]{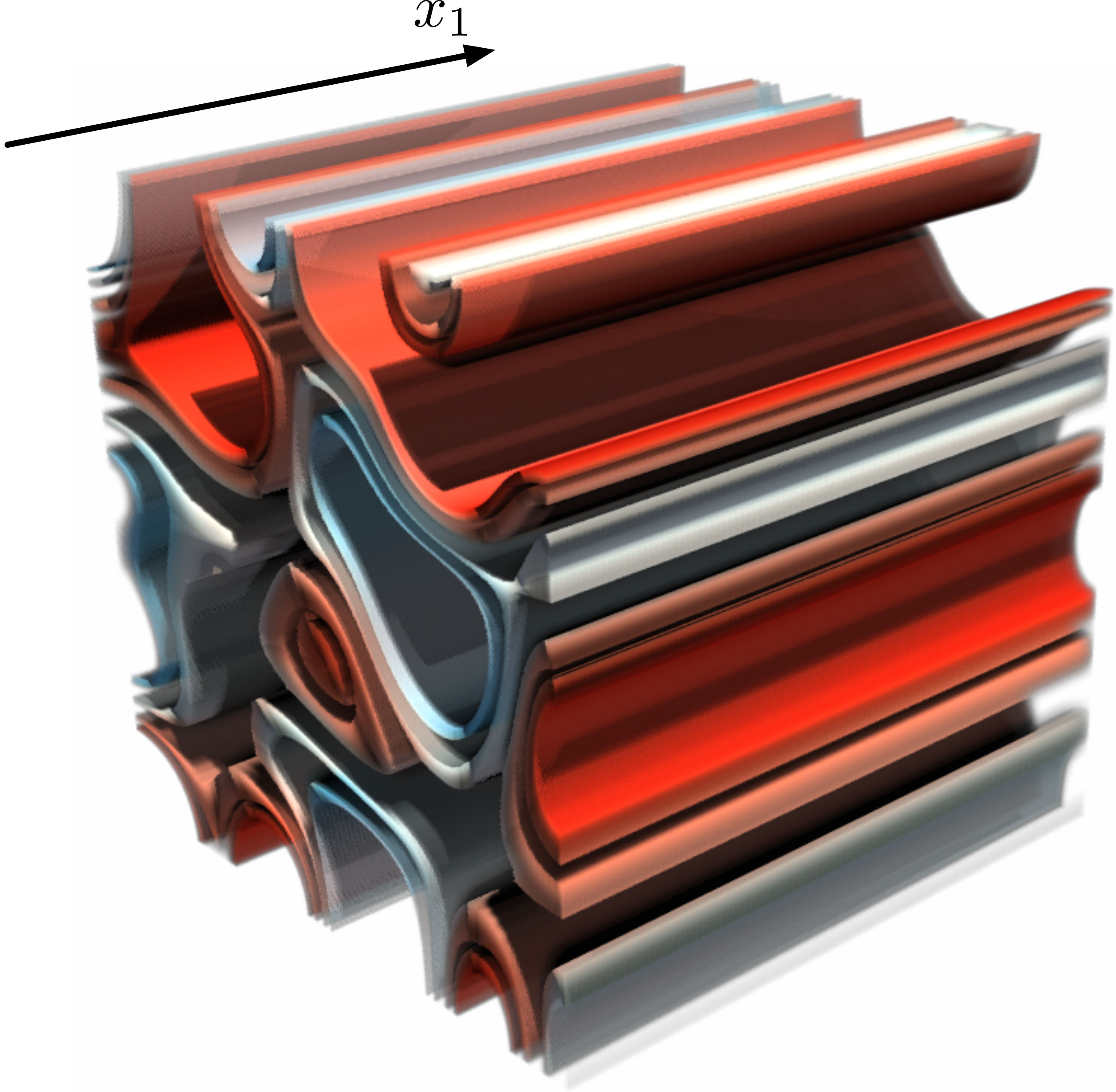}
\caption{Cross-mean of $f_{-1}$}
\end{subfigure}
\caption{Isosurface renderings for an example 3D function $f(x_1, x_2, x_3)$ (a) with target variable $x_1$. The within-mean (b) is an average over axes $x_2, x_3$ that only varies along $x_1$, whereas the cross-mean (c) is an average over axis $x_1$ that only varies along $x_2, x_3$.}
\label{fig:partials}
\end{figure}

See Fig.~\ref{fig:partials} for an example of the within- and cross-means for dimension $N = 3$ and target variable $x_1$. Note that Stages~A and~B are orthogonal to each other: the within-mean (Stage~A) is computed as an average over the non-target variables, whereas the cross-mean (Stage~B) is an average over the remaining target variables.

\begin{figure*}[h] \center
    \includegraphics[width=0.95\linewidth]{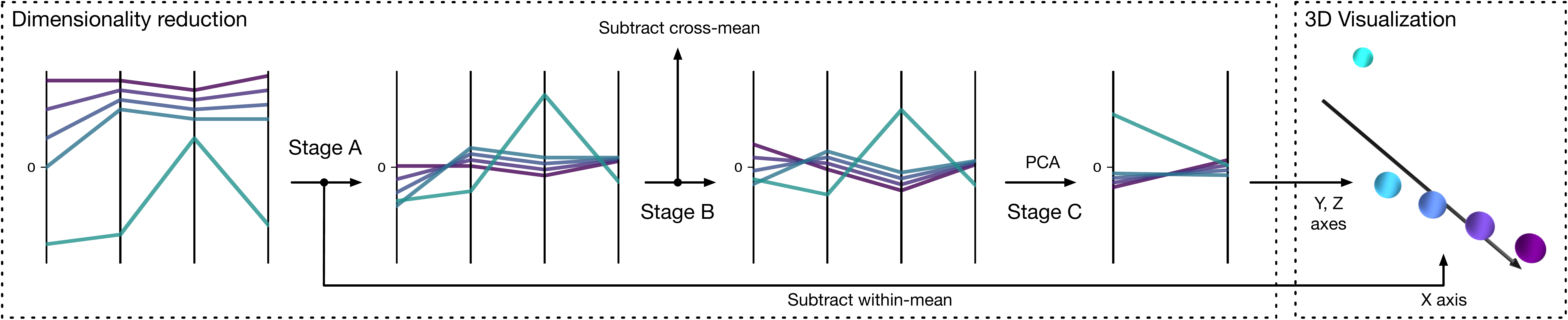}
 \caption{A schematic 4D toy example using parallel coordinates and a small set of 5 scattered points instead of multidimensional partial functions for ease of presentation. If the points are organized as rows in a $5 \times 4$ data matrix $\mat{X}$, then the within-mean $\pmb{\mu}_w$ is the row-wise mean of $\mat{X}$, whereas the cross-mean is the column-wise mean of $\mat{X}-\pmb{\mu}_w$.}
 \label{fig:parallel_coordinates}
\end{figure*}

Fig.~\ref{fig:parallel_coordinates} illustrates the full three-stage pipeline for a toy example of a 4D scattered data set displayed in parallel coordinates.

%------------------------------------------------------------------------------------------------------------------------------
\subsection{Discretization} \label{sec:discretization}

Given an infinite collection of functions, each of which is an element of an infinite-dimensional vector space, how can we find a good truncated basis for it? As a first step, let us work on a discretely sampled version of the problem, whereby we quantize the collection into a number of representative bins. This makes the procedure numerically tractable, namely via an eigensolver, so that we can approximate the original function space's KLE. Essentially, given a parameter space with $N$ inputs, w.l.o.g. we discretize the input function $f$ along each axis using $I$ bins to yield an $N$-dimensional data tensor $\set{T}$ of size $I^N$. This way, partial functions become hyperslices of tensors. Instead of a continuous 3D parameterization $(\pi_x, \pi_y, \pi_z)$, we then seek three corresponding discrete (coordinate) tensors $(\set{X}, \set{Y}, \set{Z})$, each of size $I^K$. For example, for $I = 64$ and $K=3$ we will obtain a triplet of $64^3$ coordinate volumes that can be visualized as e.g. an ensemble of $64$ 3D surfaces, each represented as a quadmesh of $64^2$ vertices. See Alg.~\ref{alg:main} for a practical, discretized version of the proposed algorithm. 

\begin{algorithm}[h]
\begin{algorithmic}[1]
\Statex 
\begin{center}\Statex \{Stage A\}\end{center}
\State \emph{// Within-mean of each partial: average over non-target variables}
\State $\set{T}_{1 \dots K} := \mathrm{mean}(\set{T}; K+1, \dots, N)$
\vspace{0.2cm}
\State \emph{// Separate and subtract the within-mean}
\State $\set{T}_{-1 \dots K} := \set{T} - \set{T}_{1 \dots K}$
\vspace{-0.2cm}
\begin{center}\Statex \{Stage B\}\end{center}
\State \emph{// Cross-mean among all partials: average over target variables}
\State $\set{T}_{\overline{-1 \dots K}} := \mathrm{mean}(\set{T}_{-1 \dots K}; 1, \dots, K)$
\vspace{0.2cm}
\State \emph{// Separate and subtract the cross-mean}
\State $\set{M} := \set{T}_{-1 \dots K} - \set{T}_{\overline{-1 \dots K}}$
\vspace{-0.2cm}
\begin{center}\Statex \{Stage C\}\end{center}
\State \emph{// Compute covariances among all pairs of tensor partials}
\State $\set{C} := \mathrm{zeros}(I, \dots, I)$ \emph{ // Tensor of size $I^{2K}$}
\For{$i_1, \dots, i_K = 1, \dots, I$}
	\For{$j_1, \dots, j_K = 1, \dots, I$}
		\State $\set{C}(i_1, \dots, i_K, j_1, \dots, j_K) = \langle \set{M}^{i_1 \cdots i_K}, \set{M}^{j_1 \cdots j_K} \rangle$
	\EndFor
\EndFor
\State $\mat{C} := \mathrm{reshape}(\set{C}, I^K \times I^K)$ \emph{ // Covariance matrix}
\vspace{0.2cm}
\State \emph{// Compute the two leading eigenpairs of $\mat{C}$}
\State $\pmb{\Lambda}, \mat{U} := \mathrm{EIG}(\mat{C}; 2)$
\vspace{0.2cm}
\State \emph{// Gather and return 3D parameterization tensors}
\State $\set{X} := \set{T}_{1 \dots K}$ \emph{ // The $x$ coordinate is exactly the within-mean}
\State $\set{Y} := \mathrm{firstColumn}(\pmb{\Lambda} \cdot \mat{U})$ \emph{ // Vector with $I^K$ elements}
\State $\set{Y} := \mathrm{reshape}(\set{Y}, I \times \dots \times I)$
\State $\set{Z} := \mathrm{secondColumn}(\pmb{\Lambda} \cdot \mat{U})$ \emph{ // Vector with $I^K$ elements}
\State $\set{Z} := \mathrm{reshape}(\set{Z}, I \times \dots \times I)$
\State \Return $\set{X}, \set{Y}, \set{Z}$
\end{algorithmic}
\caption{Input: an $N$-dimensional function $f$ discretized as a tensor $\set{T}$ of shape $I^N$ and $1 \le K < N$ variables of interest (for simplicity, here assumed to be the first $1, \dots, K$). Output: 3 tensors $\set{X}, \set{Y}$ and $\set{Z}$ that describe their discretized principal parameterization, namely a $K$-dimensional manifold in $\mathbb{R}^3$.}
\label{alg:main}
\end{algorithm}

Alg.~\ref{alg:main} relies on a range of expensive tensor operations: computing means along several axes, element-wise subtracting tensors, computing a large covariance matrix $\mat{C}$ among many tensor slices, and eigendecomposition of that matrix. In particular, the entries of $\mat{C}$ require very large-scale dot products that are non-trivial to compute. A classical method to estimate such products is Monte Carlo (MC) integration, which is simple but costly as it converges slowly~\cite{IL:15}. In addition, for higher values $K$, $\mat{C}$ may grow to become a massive dense matrix with billions of entries, so its eigendecomposition poses a challenge on its own. For example, for $K = 3$ and a moderate discretization size of 64 bins per dimension, the method must compute the leading eigenvectors of a matrix of size $64^3 \times 64^3$. While MC estimation may be sufficient in some cases for offline dimensionality reduction and visualization, it is hardly practical for interactive navigation, which is the more desirable goal. A suitable algorithm, therefore, is required.

%------------------------------------------------------------------------------------------------------------------------------
\subsection{Tensor Decomposition Algorithm}

We propose to use tensor decomposition, and in particular the tensor train (TT) model, to represent and work with our discretized parameter space. It is an extremely convenient format for the problem at hand because (a) often, it can compress a full parameter space very compactly, circumventing the curse of dimensionality; (b) allows for very fast multidimensional integration; and (c) can encode the covariance matrix needed in a TT-compressed form of its own, from which principal components are then easy to extract. The TT format approximates each entry $1 \le i_1, \dots, i_N \le I$ of our discretized tensor $\set{T}$ as a product of matrices:

\begin{equation}
\set{T}[i_1, \dots, i_N] \approx \set{T}^{(1)}[i_1] \cdot ... \cdot \set{T}^{(N)}[i_N]
\end{equation}
where every $\set{T}^{(n)}$ is a 3D tensor known as \emph{core}, namely an array of $I$ matrices indexed by $i_n$; $\set{T}^{(1)}$ and $\set{T}^{(N)}$ contain row and column vectors, respectively. In other words, the model's behavior for any dimension $n$ and any value of $i_n$ is completely governed by the elements in its corresponding matrix $\set{T}^{(n)}[i_n]$. 
The TT representation allows us to perform all required operations efficiently, provided that the input parameter space itself is given discretized and in the TT format. For instance, to compute a function's average along axes $1, ...,K$ one only needs to compute averages of the corresponding cores $\set{T}^{(1)}, \dots, \set{T}^{(K)}$. Furthermore, we do not actually require explicit expensive loops (corresponding to lines 11 to 15 in Alg.~\ref{alg:main}) to populate all elements of the $I^K \times I^K$ covariance matrix $\mat{C}$: we hold all entries of this matrix in the tensor train format and extract its leading eigenpairs efficiently in the compressed domain.

%------------------------------------------------------------------------------------------------------------------------------
\section{Geometric Interpretation}    \label{sec:properties}
%------------------------------------------------------------------------------------------------------------------------------

%------------------------------------------------------------------------------------------------------------------------------
\subsection{Approximate Isometries}

\begin{figure*}[ht]
\centering
\begin{subfigure}[b]{0.14\textwidth}
\includegraphics[width=\textwidth]{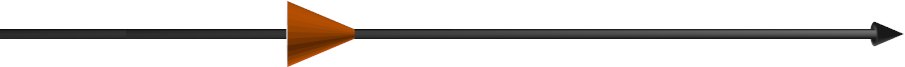} % Banzhaf, tuned
\vspace{0.76cm}
\caption{No influence}
\end{subfigure} \hfil
\begin{subfigure}[b]{0.15\textwidth}
\includegraphics[width=\textwidth]{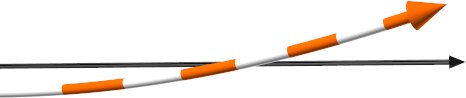}
\vspace{0.67cm}
\caption{Positive correlation}
\end{subfigure} \hfil
\begin{subfigure}[b]{0.15\textwidth}
\includegraphics[width=\textwidth]{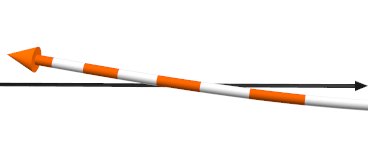}
\vspace{0.438cm}
\caption{Negative correlation}
\end{subfigure} \hfil
\begin{subfigure}[b]{0.16\textwidth}
\includegraphics[width=\textwidth]{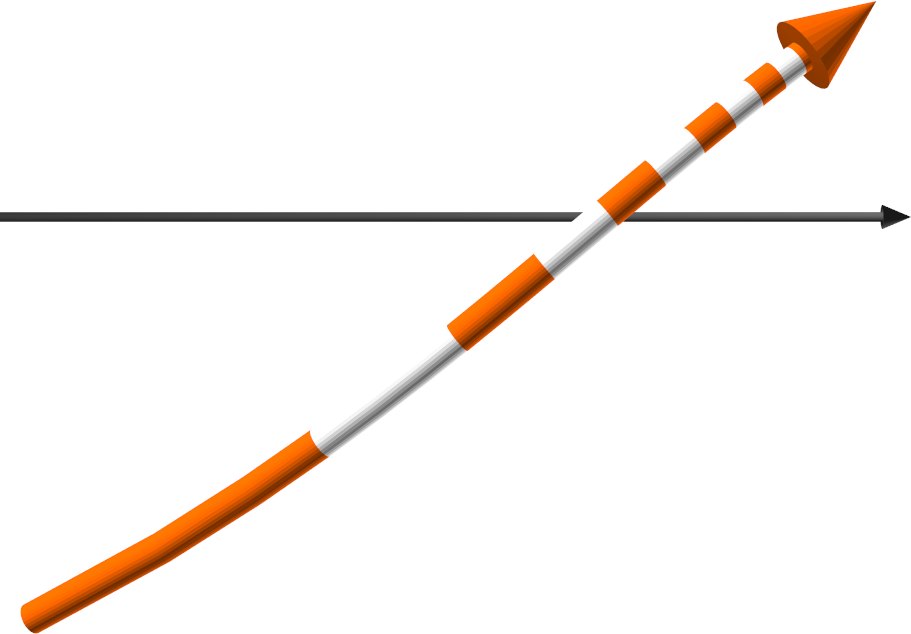} % Damped oscillator, 4, 4
\vspace{-0.45cm}
\caption{Linear deceleration}
\end{subfigure} \hfil
\begin{subfigure}[b]{0.16\textwidth}
\includegraphics[width=\textwidth]{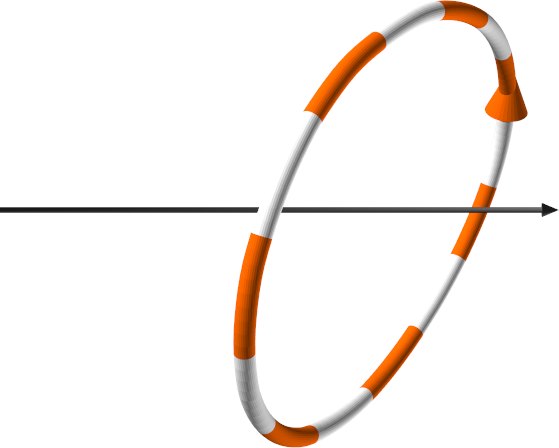} % Robot arm, 4, 4
\caption{Periodicity}
\end{subfigure} \hfil
\begin{subfigure}[b]{0.16\textwidth}
\includegraphics[width=\textwidth]{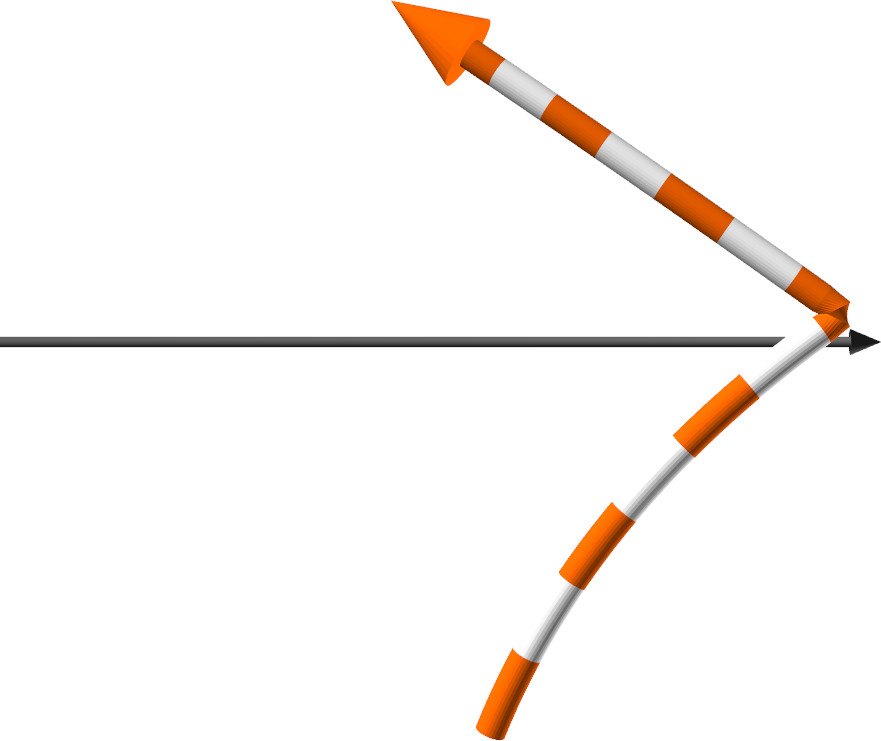}
\vspace{-0.43cm}
\caption{Bounce}
\end{subfigure}
\\
\begin{subfigure}[b]{0.16\textwidth}
\includegraphics[width=\textwidth]{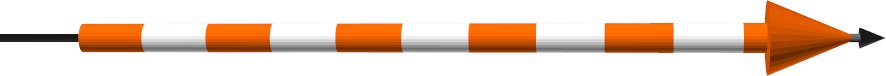}  % Damped oscillator, 7, 7
\vspace{0.82cm}
\caption{Purely additive}
\end{subfigure} \hfil
\begin{subfigure}[b]{0.16\textwidth}
\includegraphics[width=0.9\textwidth]{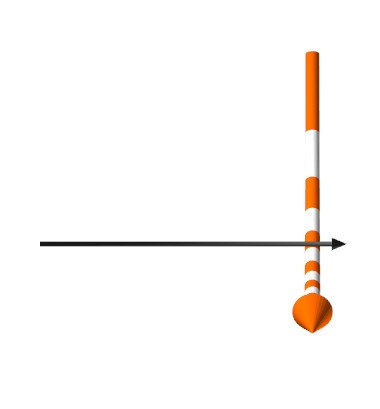}
\vspace{0.16cm}
\caption{Purely high-order}
\end{subfigure} \hfil
\begin{subfigure}[b]{0.16\textwidth}
\includegraphics[width=\textwidth]{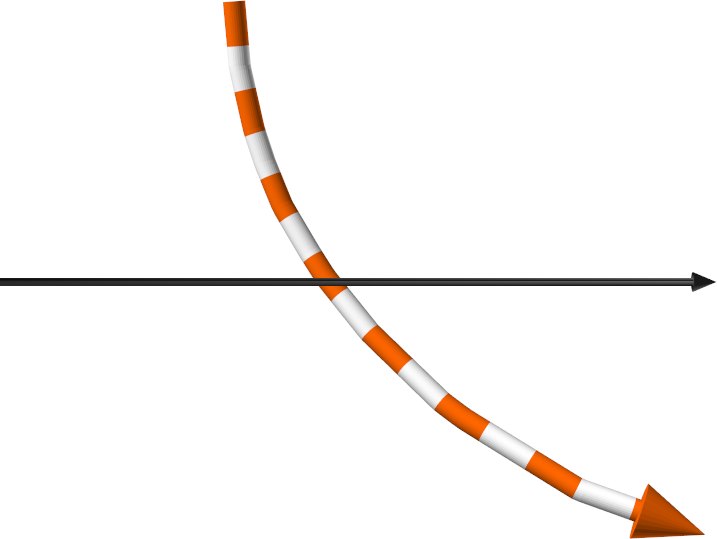} % Robot arm, 7, 7
\vspace{-0.38cm}
\caption{Changing from high-order to more additive behavior}
\end{subfigure} \hfil
\begin{subfigure}[b]{0.16\textwidth}
\includegraphics[width=\textwidth]{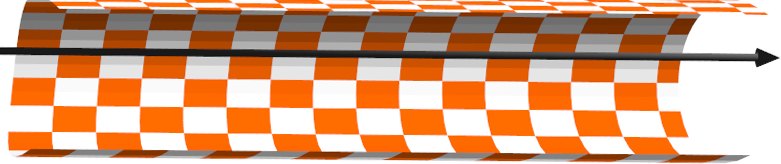}
\vspace{0.22cm}
\caption{Additive + higher-order, little interaction}
\end{subfigure} \hfil
\begin{subfigure}[b]{0.16\textwidth}
\includegraphics[width=\textwidth]{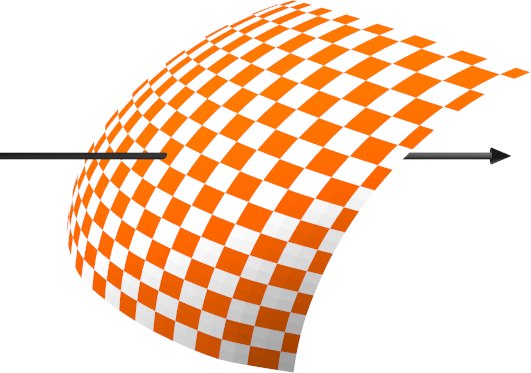} % Robot arm, 1, 3
\vspace{-0.25cm}
\caption{Mixed effects}
\end{subfigure} \hfil
\begin{subfigure}[b]{0.16\textwidth}
\includegraphics[width=\textwidth]{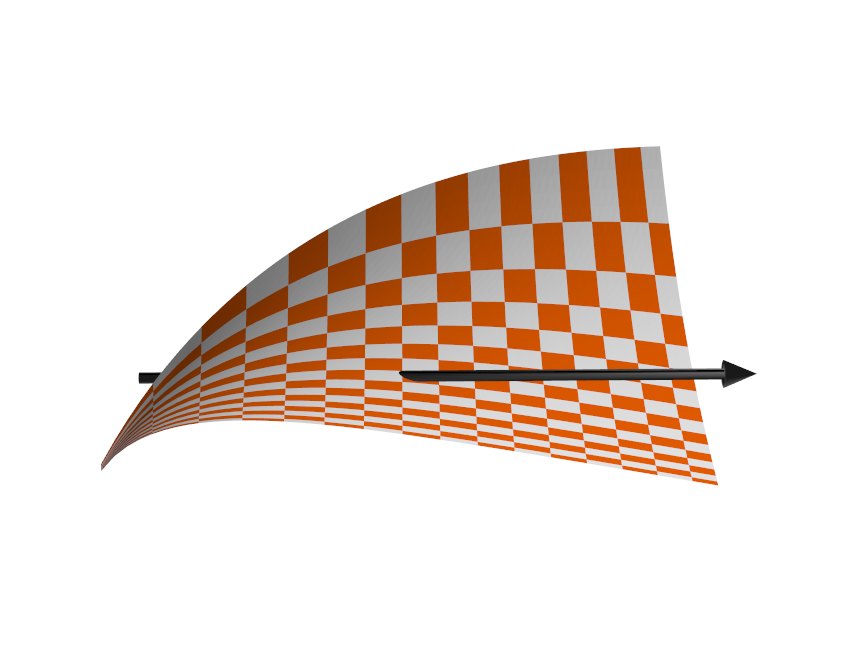}
\vspace{-0.02cm}
\caption{Very strong interaction}
\end{subfigure}
\caption{The proposed 1D and 2D principal parameterizations capture a wide range of single- and multiple-effect patterns as well as global/local properties. For example, (f) shows a non-differentiable point due to a $\min$ function, whereas (l) demonstrates a variable that is not influential when another variable is small, but is quite influential otherwise. The $x$-axis is depicted in each case as a black arrow and marks the model's average over all abstracted variables for any specific values of the parameterized target variables (color arrows and surfaces). Rows and columns in the surfaces' checkerboard patterns reflect variable isolines, i.e. the path followed when a variable moves and the other one is fixed.}
\label{fig:examples}
\end{figure*}

\begin{figure}[h!]
\centering
\begin{subfigure}[c]{0.54\columnwidth}
\includegraphics[width=1\columnwidth]{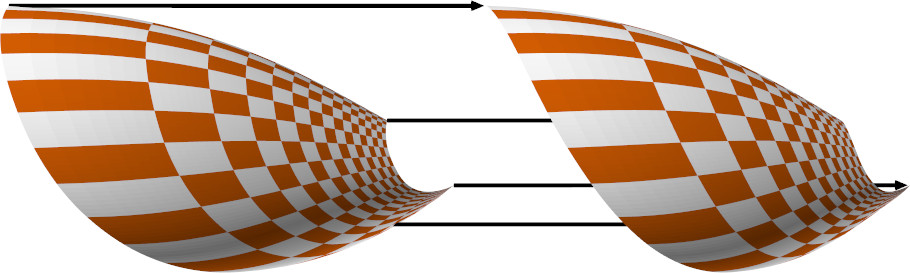}
\caption{Linear effect}
\end{subfigure}
\hfil
\begin{subfigure}[c]{0.34\columnwidth}
\includegraphics[width=0.9\columnwidth]{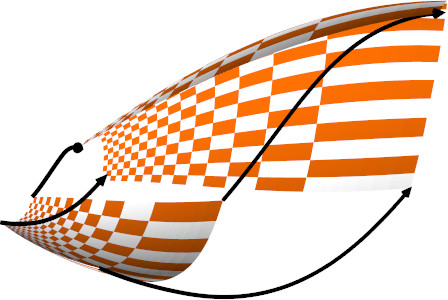}
\caption{Non-linear effect}
\end{subfigure}
\caption{We can display 3D trajectory curves so as to track individual points in a parameterized surface as a third variable moves.}
\label{fig:tracking}
\end{figure}

The truncated PCA yields the projection $\pi$ that, by means of a reduced orthonormal basis, best preserves a given collection of vectors in the $L^2$ sense: 

\begin{equation*}
\argmin_{\pi} \sum_{\vec{u}} \|\vec{u} - \pi^{-1}(\pi(\vec{u}))\|^2.
\end{equation*}
where $\pi^{-1}(\cdot)$ is the expansion back into the original high-dimensional space using the same basis. This means that distances between vectors are also preserved well:
\begin{equation*}
\|\vec{v} - \vec{u}\| \approx \|\pi(\vec{v}) - \pi(\vec{u})\|,
\end{equation*}
and likewise relative distance changes:
\begin{equation*}
\|\vec{w} - \vec{v}\| - \|\vec{v} - \vec{u}\| \approx \|\pi(\vec{w}) - \pi(\vec{v})\| - \|\pi(\vec{v}) - \pi(\vec{u})\|,
\end{equation*}
and similarly for any level of repeated subtraction. In other words, notions like \emph{speed of change} or \emph{acceleration} tend to be reflected well in the projected space. This has important and desirable consequences from a visualization point of view. So, for example, if a collection of partial functions $f^{x_n}$ for $0 \le x_n \le 1$ follows a straight line in $\set{F}$, then its projection $\pi:[0, 1] \to \mathbb{R}^3$ will necessarily evolve in a linear fashion:
\begin{equation*}
\pi(x_n) = \pi(0) + x_n \cdot (\pi(x_n) - \pi(0)),\ 0 \le x_n \le 1
\end{equation*}
which is a straight line connecting 3D points $\pi(0)$ and $\pi(1)$. Furthermore, a curved line hints at a sequence of vectors that changes non-linearly. Sudden changes in the curve mirror sudden changes also in the original high-dimensional $\set{F}$ (as intuitively expected), periodic behavior is mapped to rings, etc. Also, note that the global mean of the model $\mathbb{E}[f]$ coincides with the barycenter of any principal parameterization in 3D, which has the form $(\mathbb{E}[f], 0, 0)$. The cosine similarity $\frac{\vec{u} \cdot \vec{v}}{\|\vec{u}\| \cdot \|\vec{v}\|}$ is approximately preserved as well, and is rendered in 3D as angles between vectors. To help gain intuition and demonstrate the expressive power of the proposed parameterizations, we show a number of examples in Figs.~\ref{fig:examples} and~\ref{fig:tracking}. All were taken from the models listed later in Sec.~\ref{sec:models}.

Certainly, since we are projecting vectors of huge dimensionality using three basis elements only, much of the detail along the unimportant variables is likely to be smoothened out. On the other hand, because the manifold has as many dimensions as there are variables of interest, the trajectories of these variables are captured well and can be tracked visually in full detail. For example, any sharp corner or feature in a curve traces back unambiguously to one specific value of its variable $x_n$. We argue this is a strength of the proposed method: it is able to abstract complex spaces over many dimensions while still retaining full resolution along a few selected target variables.

%------------------------------------------------------------------------------------------------------------------------------
\subsection{Global Sensitivity Analysis} \label{sec:sobol_interpretation}

As outlined in the introduction, there are several interesting connections relating our projection $\pi$ with the Sobol indices~\cite{Sobol:90} (Sec.~\ref{sec:related_work}). Recall that for the Sobol index of the $n$-th variable we have $S_n \propto \mathrm{Var}[f_n]$, whereas the total Sobol index $S^T_n$ accounts for both the additive and high-order effects: $S^T_n \propto \mathrm{Var}[f_n] + \mathrm{Var}[f_{-n}]$. Furthermore, we have that $\mathrm{Var}[f_n] = \|f_n\|^2 \propto \|\pi_x\|^2$ (exact projection on the $x$-axis) and $\mathrm{Var}[f_{-n}] = \|f_{-n}\|^2 \propto \|(\pi_y, \pi_z)\|^2$ (approximately; it is the projection on the $yz$-plane using a truncated expansion). Therefore,

\begin{itemize}
	\item The curve's evolution along the $x$-axis mirrors the corresponding partial's mean value $\mathbb{E}[f^{x_n}]$ as $x_n$ moves, and $S_n$ is proportional to the curve's variance along that axis. In other words, by tracking the curve's $\pi_x$ coordinate we can infer the overall additive behavior of our $N$-dimensional model as that variable progresses. In particular, the correlation between $x_n$ and the model output equals that between $x_n$ and the curve's $\pi_x$: $\rho(x_n, f) = \rho(x_n, \mathbb{E}[f^{x_n}]) = \rho(x_n, \pi_x(x_n))$. Thus, curves that generally move towards the right of the $x$-axis indicate a positive correlation, and vice versa. Any purely additive variable $x_n$ (i.e. $S_n = S^T_n$) will not use the $yz$-plane, i.e. is mapped to a line segment that is perfectly aligned to the $x$-axis. See also examples in Fig.~\ref{fig:examples}a,b,c,d,g.
	\item The higher-order component measures exclusively the influence due to the interplay between the variable of interest and the rest of variables. This interaction is reflected as variations along the $yz$-plane. The manifold's second-order moment on that plane, i.e. its summed squared distance to the $x$-axis, is proportional to $S^T_n - S_n$. Any purely high-order variable $x_n$ (i.e. that \emph{does not} influence the model's average, $S_n = 0$) does not move along the $x$-axis, but only on the $yz$-plane, e.g. as in Fig.~\ref{fig:examples}h.
	\item The total index $S^T_n$ measures both effects and is approximately proportional to the total second spatial moment (that is, including all $X$, $Y$, and $Z$ axes) of the principal parameterization. In other words, the more \emph{spread out} the parameterization is, the more global influence its variable has on the model. Conversely, a tuple of irrelevant variables will be collapsed into a point.
\end{itemize}

In a nutshell, additive effects make the curve move \emph{along} the $x$-axis, while high-order interactions \emph{pull} it away in various ways.

%------------------------------------------------------------------------------------------------------------------------------
\subsection{Local Sensitivity Analysis} \label{sec:local_sensitivity}

\begin{figure}[h!] % local_behavior.py
\centering
\begin{subfigure}[b]{0.4\columnwidth}
\includegraphics[width=\columnwidth]{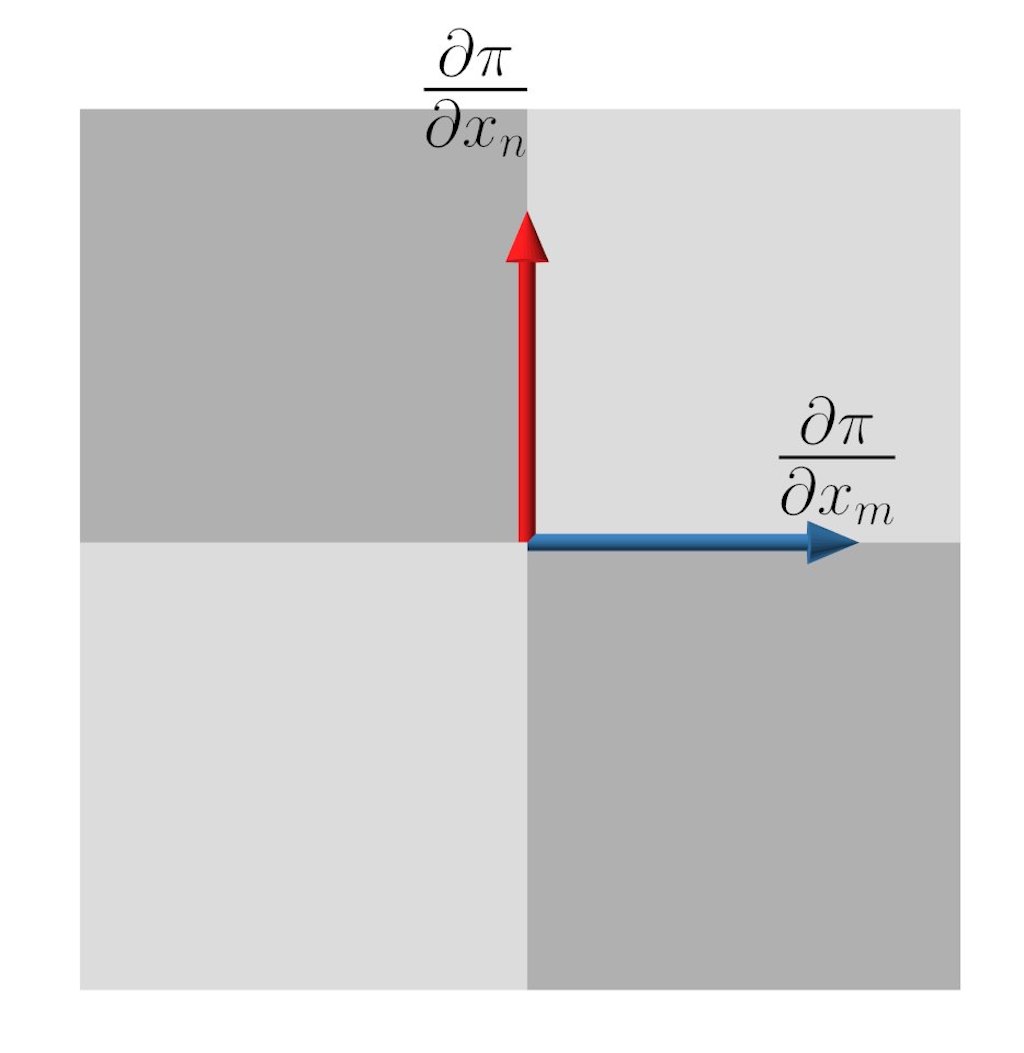}
\caption*{$\frac{\partial \pi}{\partial x_n} \cdot \frac{\partial \pi}{\partial x_m} \approx 0$}
\caption{Non-aligned derivatives}
\end{subfigure} \hfil
\begin{subfigure}[b]{0.45\columnwidth}
\includegraphics[width=\columnwidth]{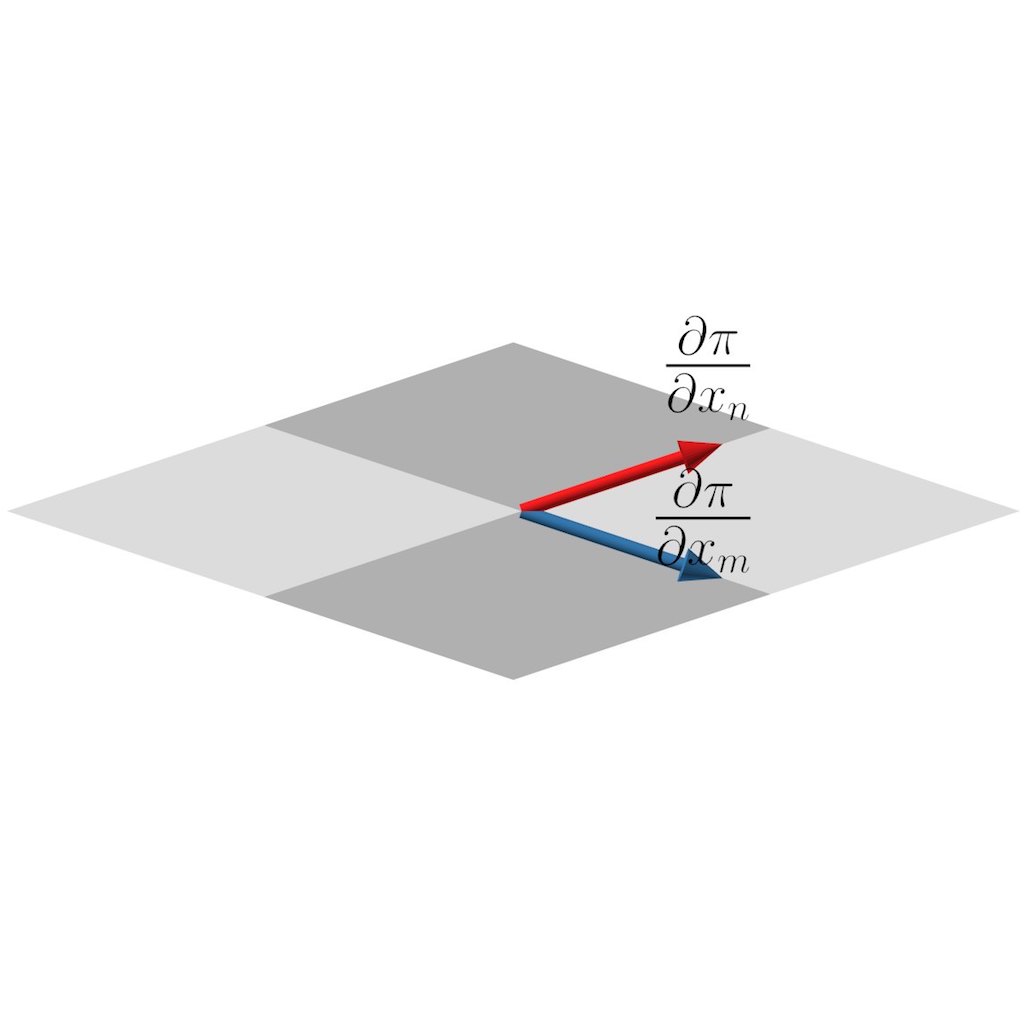}
\vspace{-0.9cm}
\caption*{$\frac{\partial \pi}{\partial x_n} \cdot \frac{\partial \pi}{\partial x_m} \approx \|\frac{\partial \pi}{\partial x_n}\| \cdot \|\frac{\partial \pi}{\partial x_m}\|$}
\caption{Well-aligned derivatives}
\end{subfigure}
\\
\vspace{0.3cm}
\hspace{0.25cm}
\begin{subfigure}[b]{0.31\columnwidth}
\includegraphics[width=\columnwidth]{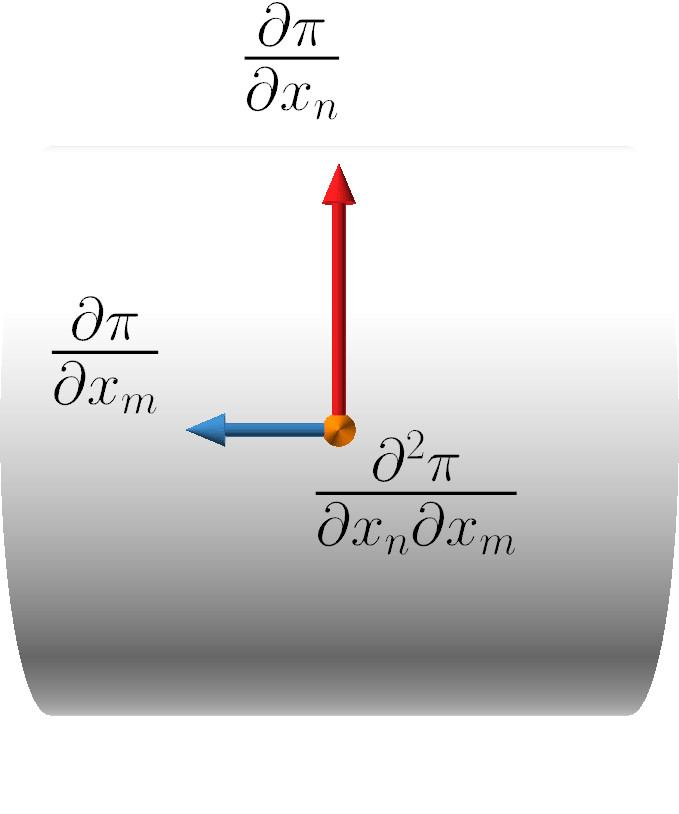}
\caption*{$\frac{\partial^2 \pi}{\partial x_n \partial x_m} = 0$}
\caption{No joint interaction}
\end{subfigure} \hfil
\begin{subfigure}[b]{0.45\columnwidth}
\includegraphics[width=\columnwidth]{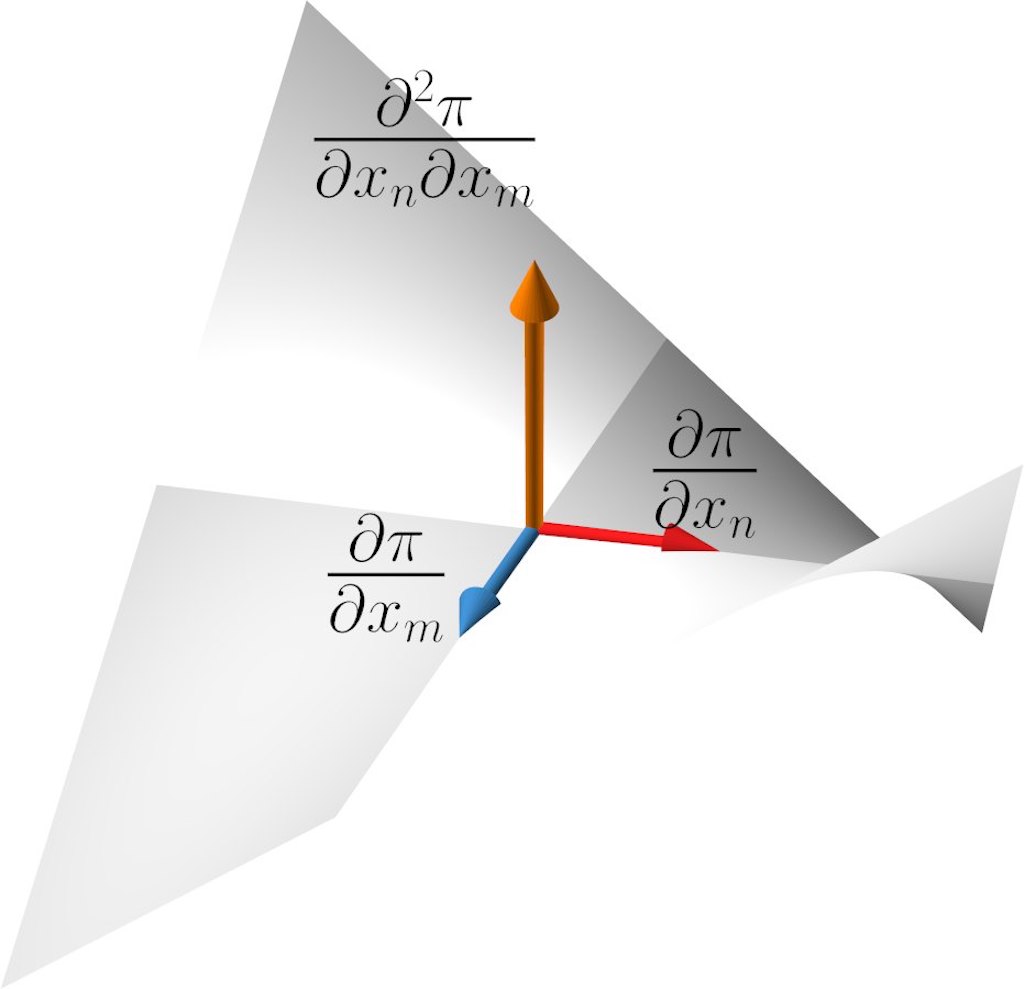}
\vspace{-0.65cm}
\caption*{$\frac{\partial^2 \pi}{\partial x_n \partial x_m} \ne 0$}
\caption{Strong joint interaction}
\end{subfigure}
\caption{Different types of pair-wise effects and interactions at the local level are rendered as differential features on a principal surface.}
\label{fig:surfaces}
\end{figure}

The differential structure around a point of the manifold tells us about the local behavior of the variables of interest when they take some specific value. Consider the derivatives $\frac{\partial \pi}{\partial x_n}$ and $\frac{\partial \pi}{\partial x_m}$ of a principal surface at a certain point $x_n, x_m$. The angle between them is a proxy of the similarity between the local effects caused by small increments of the $n$-th and the $m$-th variables on the high-dimensional model. For example, if both exert an identical influence on $f$ around that point's neighborhood, their derivative vectors will be aligned and equal.
The mixed derivative $\frac{\partial^2 \pi}{\partial x_n \partial x_m}$ measures the joint interaction between those variables in the Sobol sense, i.e. their influence on the model that is not due to the sum of individual additive changes of $x_n$ and $x_m$ in the neighborhood. See Fig.~\ref{fig:surfaces} for some illustrations.

\begin{figure}[h!]
\centering
\begin{subfigure}[c]{0.45\columnwidth}
\includegraphics[width=\columnwidth]{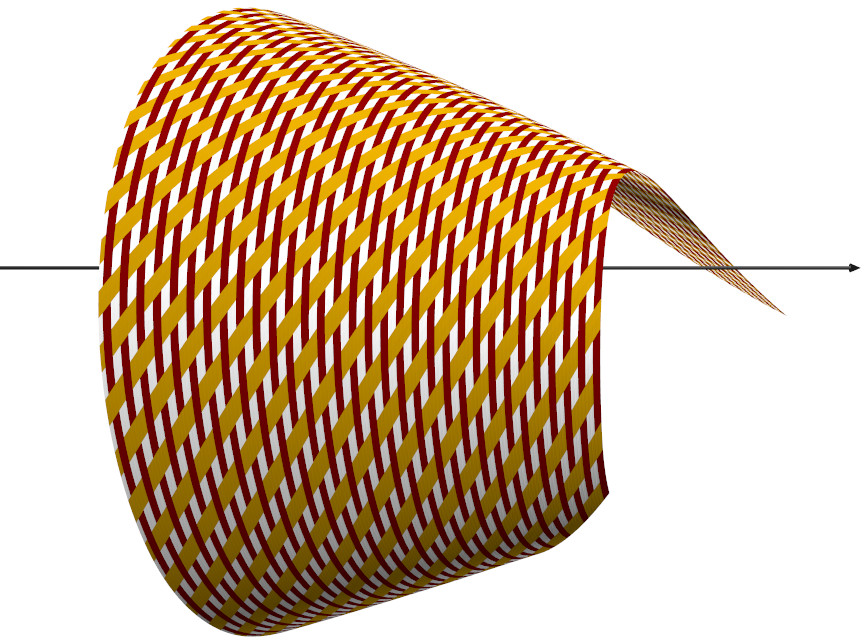}
\vspace{-0.225cm}
\caption{Isolines}
\end{subfigure} \hfil
\begin{subfigure}[c]{0.45\columnwidth}
\includegraphics[width=\columnwidth]{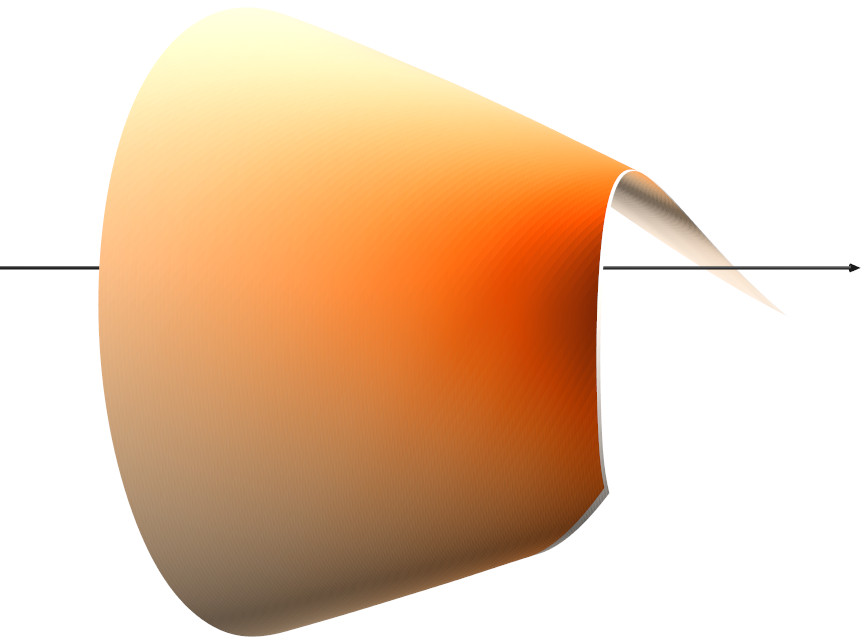}
\caption{$\left \lVert \frac{\partial^2 \pi}{\partial x_n \partial x_m} \right \rVert$}
\end{subfigure} \\ \vspace{0.2cm}
\begin{subfigure}[c]{0.45\columnwidth}
\includegraphics[width=\columnwidth]{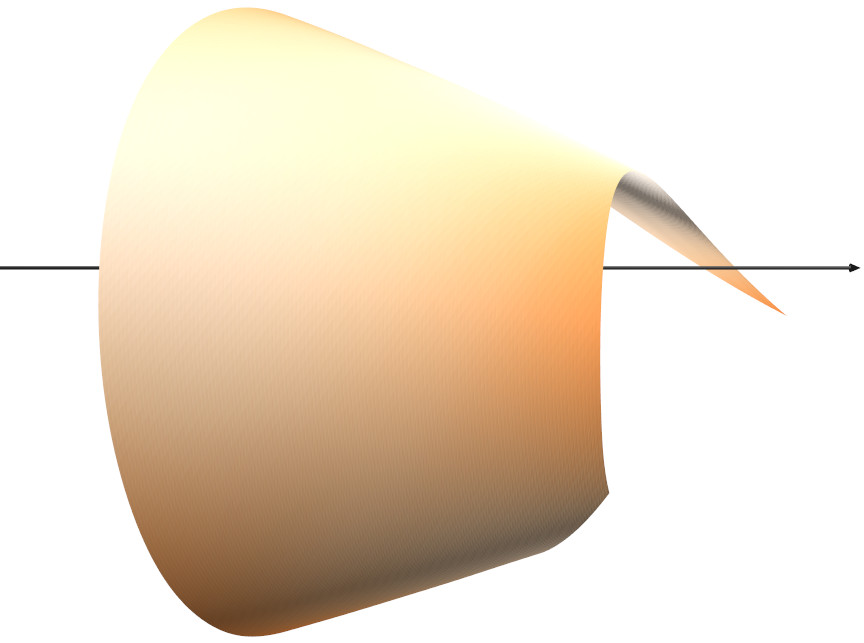}
\vspace{-0.36cm}
\caption{$\eps(x_n, x_m)$}
\end{subfigure} \hfil
\begin{subfigure}[c]{0.45\columnwidth}
\includegraphics[width=\columnwidth]{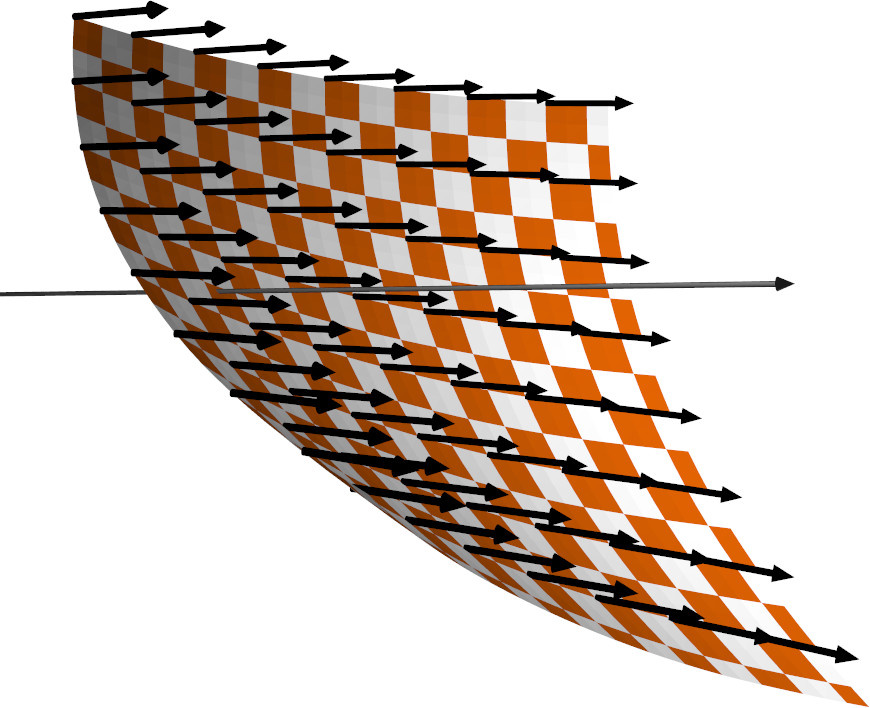}
\caption{$\frac{\partial \pi}{\partial x_k}, k \ne n, m$}
\end{subfigure}
\caption{We can convey different types of local information (see also Sec.~\ref{sec:local_sensitivity}) via surface texturization: (a) interwoven parameter isolines; (b) magnitude of local pair-wise interactions, computed as the mixed derivative's norm at each point; (c) PCA projection error at each point; d) vector field of derivatives w.r.t. a third variable.}
\label{fig:colorizations}
\end{figure}

Alternatively, we can also compute the projection error for each point as the distance between the corresponding partial and its approximation $\eps(x_n, x_m) := \|f^{x_n, x_m} - \pi^{-1}(\pi(f^{x_n, x_m}))\|$. Such local scalar properties can effectively be displayed as a color map texture on the principal surfaces; see a range of examples in Fig.~\ref{fig:colorizations}.

%------------------------------------------------------------------------------------------------------------------------------
\section{Results} \label{sec:results}
%------------------------------------------------------------------------------------------------------------------------------

We used Python 3.5, Qt and OpenGL to implement the proposed algorithm\footnote{Our code is available in \url{https://github.com/rballester/ttpca}.}. All visual results are displayed using a range of custom widgets and diagrams. Our visualization front-end uses \emph{PyQtGraph}~\cite{pyqtgraph}, a 2D/3D graphics and GUI library for scientific Python, as well as \emph{PyQt} and \emph{PyOpenGL} which enable Python interfacing with Qt and OpenGL, respectively. Besides NumPy, our numerical back-end exploits the auxiliary libraries \emph{ttpy}~\cite{ttpy} and \emph{ttrecipes}~\cite{ttrecipes} for tensor train manipulation.

%------------------------------------------------------------------------------------------------------------------------------
\subsection{Models Tested} \label{sec:models}

We considered five analytical models of varying complexity and dimensionality:

\begin{itemize}
	\item The \emph{Nassau County}~\cite{Banzhaf:65}: a 6D voting system where any coalition of political agents can pass a motion if and only if their combined votes reach a simple majority. It is a naturally discrete problem; we model all 64 possible coalitions as a small tensor of size $2^6$.
	\item The \emph{Cell Cycle}~\cite{GDC:98}: a 4D multivalued time-dependent model, which builds on an earlier model by Goldbeter~\cite{Goldbeter:91} and is discretized as a tensor of size $32 \times 32 \times 32 \times 64 \times 5$. The second-to-last dimension is the time $t$, while the last dimension indexes the five outputs of the model.
	\item The \emph{Damped Oscillator}~\cite{DSD:13}: an 8D physical model measuring the peak force in a spring system connecting two oscillating masses. This and the following models were discretized using 64 bins per dimension.
	\item The \emph{Robot Arm}~\cite{AO:01}: an 8D model measuring the distance attained by a 4-segment robot arm. It includes an irrelevant variable and three periodic variables (elbow joint angles moving between $0$ and $2 \pi$)
	\item The \emph{Ebola Spread}~\cite{DCKJP:18}: an 8D model predicting Ebola virus infection rate in Liberia and Sierra Leone, based on statistics from the 2014 outbreak. Two variables can be influenced to decrease the number of infections by allocating resources, whereas the rest mostly depend on environmental factors.
\end{itemize}

We built TT metamodels for those use cases by means of the \emph{cross-approximation} adaptive sampling algorithm~\cite{OT:10}, in particular the implementation released in the \emph{ttpy} Python toolbox~\cite{ttpy}. It is an incremental sampling technique that uses a growing validation set $\set{X}$ at each sample acquisition step; the process is stopped as soon as the relative error w.r.t. the current prediction $\tilde{\set{X}}$ is below a user-defined threshold: $\|\tilde{\set{X}} - \set{X}\| / \|\set{X}\| \le \eps$. We used $\eps := 10^{-4}$ in all cases.

%------------------------------------------------------------------------------------------------------------------------------
\subsection{Curve Array} \label{sec:curve_array}

The simplest kind of diagram we implemented is a collection of $N$ static parameterized curves in 3D, one per input variable $x_1, \dots, x_N$. This basic visualization conveys the summarized global structure of the model as its inputs move individually: among others, it reflects correlations between the model's output and each of its variables. Note that inter-variable interactions are only given in abstracted form as curve movements on the $yz$-plane.

\begin{figure}[h] \center
    \includegraphics[width=1\linewidth]{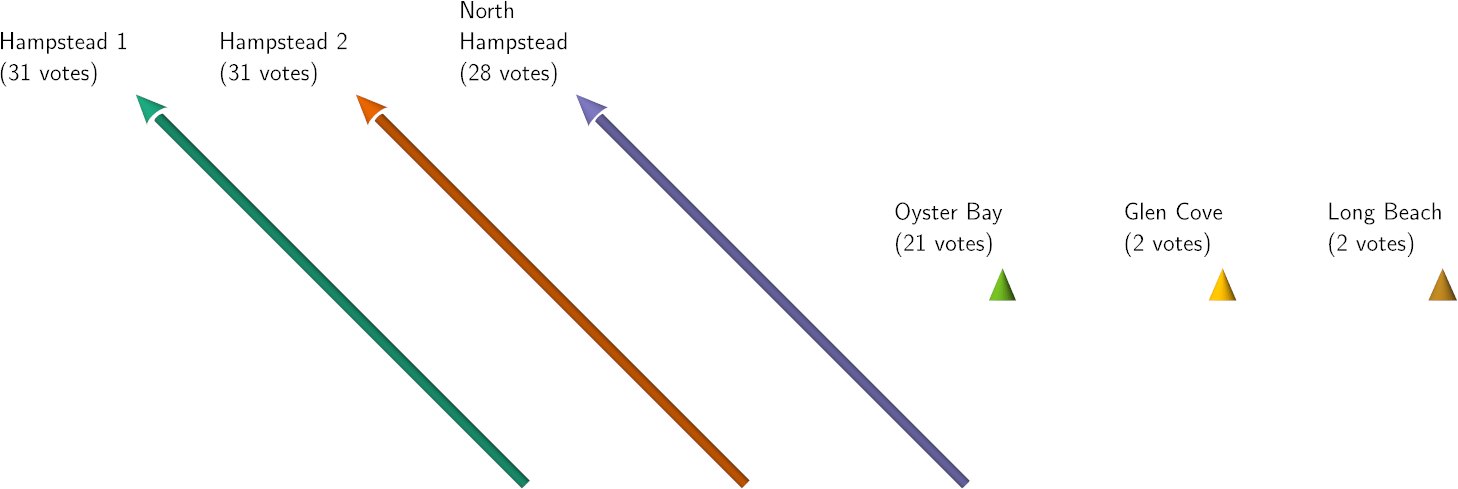}
 \caption{When applied to a majority voting system, our parameterization yields the Banzhaf power index~\cite{Banzhaf:65}. Depicted is Banzhaf's original example for the 1964 Nassau County Board of Supervisors. He argued that half of the districts were actually powerless even though they collectively held more than 1/5th of the total votes.}
 \label{fig:banzhaf}
\end{figure}

Let us start with the small \emph{Nassau County} example in order to gain intuition on the proposed method of operation. This model considers 6 county districts with different voting weights. In order for any political motion to succeed, it must be backed by a coalition of districts that reach a vote majority. In political sciences and game theory, the \emph{Banzhaf power index}~\cite{Banzhaf:65} is often used for this kind of settings to assess the true influence of individual agents (which may be far from their nominal voting weight). For each agent, its index is defined as the fraction of all possible winning coalitions in which it is a necessary member, i.e. the coalition reaches a majority but would not do so without that agent.
Since each district party can either be or not be in a coalition, we model the problem using a binary variable for each. The domain is thus $\Omega = \{0, 1\}^6$. We arrange all possible coalitions in a small tensor of size $2^6 = 64$, where each entry is 1 if the corresponding coalition would reach majority and 0 otherwise. We then compute an array of principal curves using our method. For each district, its curve has only two points, and is thus a line segment that connects them. We have found the variance of these two endpoints (i.e. the segment's squared length) to be proportional to their corresponding district's Banzhaf power index; see also Fig.~\ref{fig:banzhaf}.

\begin{figure}[h]
\centering
\begin{subfigure}[c]{1\columnwidth}
\includegraphics[width=\textwidth]{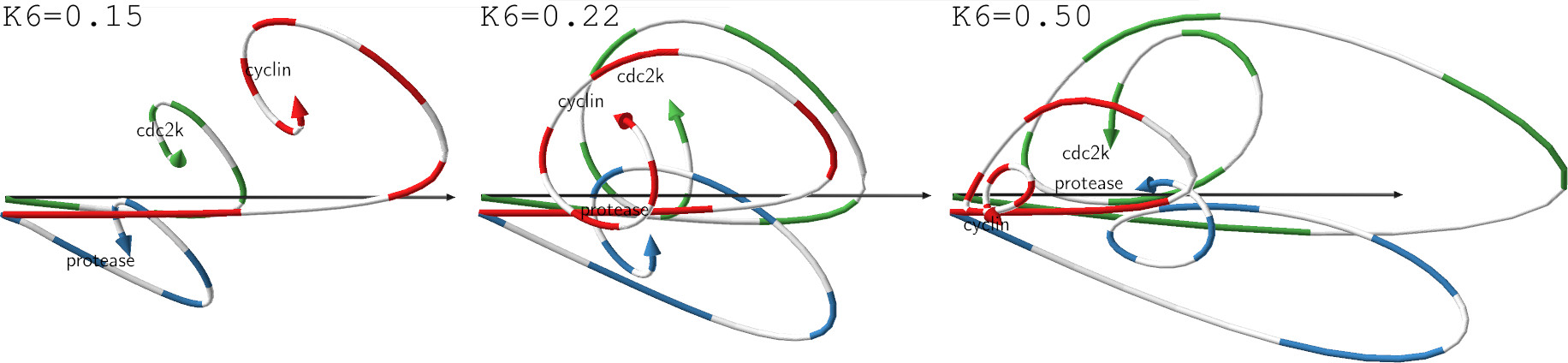}
\caption{Oscillating concentrations}
\label{fig:multioutputs_oscillating}
\end{subfigure}
\\
\vspace{0.25cm}
\begin{subfigure}[c]{1\columnwidth}
\includegraphics[width=\textwidth]{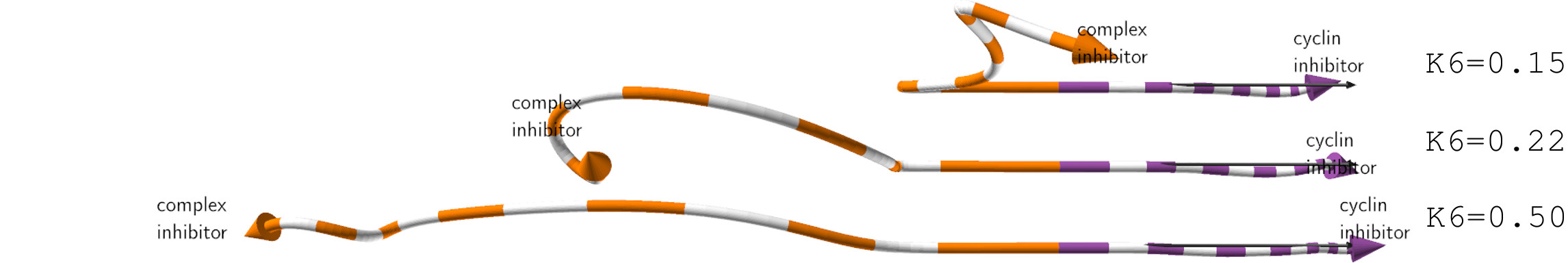}
\caption{Inhibitors}
\end{subfigure}
\caption{A dynamic 5-valued ODE modeling the cell division cycle~\cite{GDC:98} and its 5 principal curves (one per output quantity). They are shown in two group plots of 3 and 2 curves and variable of interest $t$ (time), with increasing values of the input parameter \texttt{K6}.}
\label{fig:multioutputs}
\end{figure}

As a second, more complex example we consider a 4D, 5-valued system of ordinary differential equations (ODE), namely the \emph{Cell Cycle}. It is a time-dependent system modeling protein concentrations during cell division, hence the temporal axis $t$ is particularly important as an explanatory variable. We select $t$ as our variable of interest and gather all 5 outputs of the ODE into an extra dimension for a joint analysis. This way, we enforce their 5 principal curves to share the same 3D system of projected coordinates. We have furthermore added a slider to govern any of the 4 parameters separately and thus add one extra degree of user interaction. The slider can be adjusted in real time and prompts an immediate update on the 5 curves displayed. Parameter \texttt{K6} was found to have a strong effect on the speed of change of several outputs; see Fig.~\ref{fig:multioutputs} for some example renderings.

Note that several quantities (Fig.~\ref{fig:multioutputs_oscillating}) exhibit an oscillatory pattern as is explained by their interaction with each other in the ODE. They differ in their asymptotic behavior: they oscillate around different concentration levels, and the attractor point is in each case differently affected by \texttt{K6}. Besides compactly showing the rate of evolution of each quantity as time progresses, the proposed visualization also reflects periods of similarity between two or more curves' behaviors.

%------------------------------------------------------------------------------------------------------------------------------
\subsection{Plot Matrix} \label{sec:plot_matrix}

\begin{figure}[h!] \center
    \includegraphics[width=0.7\linewidth]{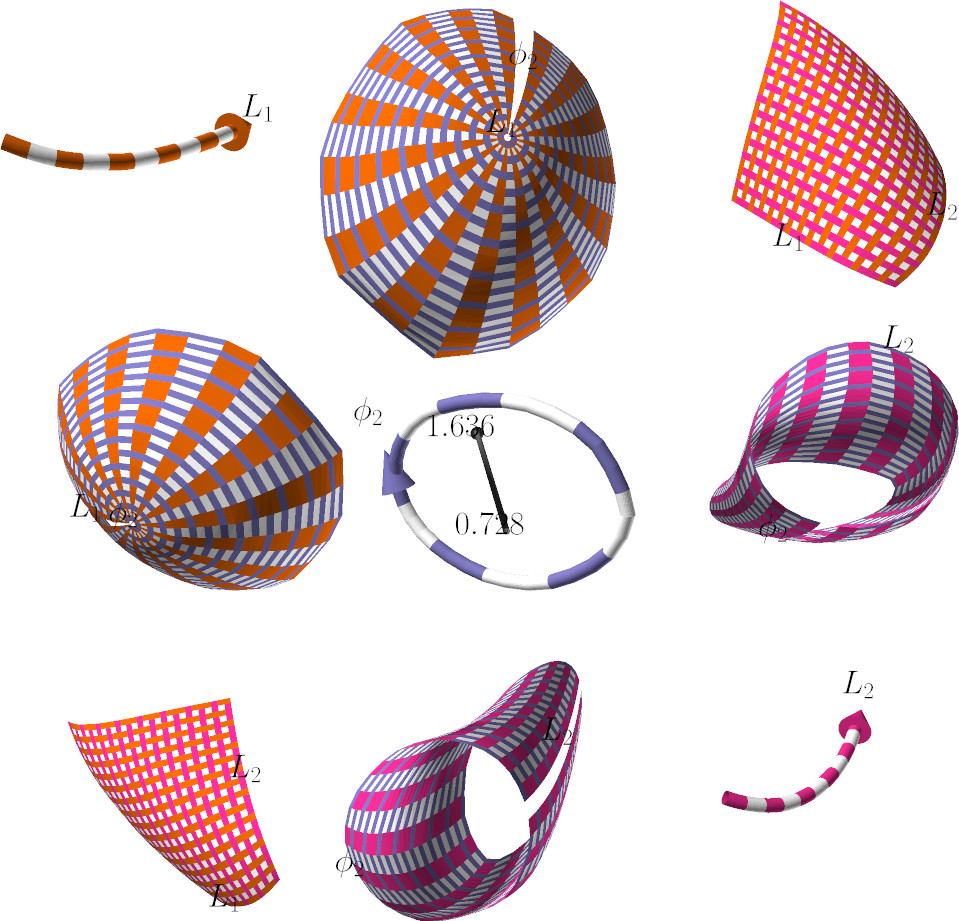}
 \caption{Plot matrix diagram (zoomed-in version showing 3 variables only) giving a compact depiction of single and pair-wise effects in the \emph{Robot Arm} high-dimensional model. The user can navigate freely within the 3D scene in order to magnify and observe details from any desired angle. See the paper teaser for zoomed-out examples.}
 \label{fig:hyperslice}
\end{figure}

The SPLOM (scatterplot matrix) and the HyperSlice~\cite{WL:93} are diagrams that arrange pair-wise relations in a square matrix fashion and line up unary items along its diagonal. Drawing on this idea we combine all 1st and 2nd-order parameterizations in an $N \times N$ plot matrix where every entry $(n, m)$ contains the principal surface for variables $x_n$ and $x_m$. The special case $n = m$ yields a curve. This diagram generalizes and is more expressive than the curve array; see Fig.~\ref{fig:hyperslice} and the teaser Fig.~\ref{fig:teaser} for some examples using the \emph{Damped Oscillator} and \emph{Robot Arm} models. For example, the periodic variable $\phi_2$ (corresponding to the angle in a robot elbow) stands out as the purple central curve in Fig.~\ref{fig:hyperslice}. Its interaction with non-periodic variables (lengths $L_1$ and $L_2$ of that elbow's arm segments) is encoded as the purple-orange and purple-magenta umbrella- and cylinder-like parameterized surfaces.

%------------------------------------------------------------------------------------------------------------------------------
\subsection{Variable Selection}

\begin{figure}[h] \center
 \includegraphics[width=0.5\linewidth]{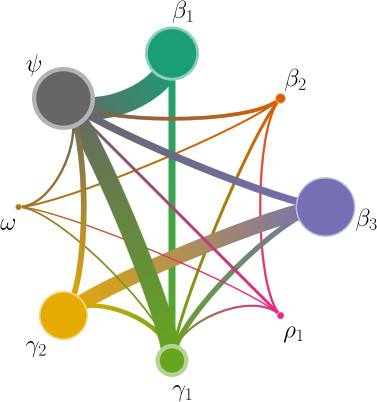}
 \caption{Our contextual radial graph (\emph{Ebola Spread} model) displays all 1st- and 2nd-order Sobol indices for any given values of the current active tuple of variables. Additive and high-order effects are shown as darker and lighter colors, respectively.}
 \label{fig:sobol_minimap}
\end{figure}

The plot matrix layout outlined above is highly compact, but it becomes impractical for more than 2 variables of interest as the number of possible combinations increases exponentially. For this reason, we support various forms of \emph{variable selection}.
There is a tight association between our diagrams and various Sobol indices as discussed in Sec.~\ref{sec:sobol_interpretation}. This gives us a forthright criterion to select interesting variables interactively: high Sobol indices reveal strong effects and interactions, and vice versa. Furthermore, such indices can be extracted efficiently from any TT surrogate~\cite{DKLM:14, BPP:17}. We have implemented a contextual minimap as shown in Fig.~\ref{fig:sobol_minimap} that extends the \emph{fanovaGraph} sensitivity analysis chart~\cite{FRM:13}. It is a graph in a circular layout that displays all first- and second-order Sobol indices of either the whole function or any arbitrary partial function. We furthermore use two palettes, one for darker and one for lighter colors. The area of the darker inner circle within each variable's node is proportional to its additive effect, while the lighter outer circle encodes its higher-order effect. The same applies to arcs for order-2 indices using the width instead of area. 

%------------------------------------------------------------------------------------------------------------------------------
\subsection{Widget-based Tool}

\begin{figure}[h] \center
\includegraphics[width=1\columnwidth]{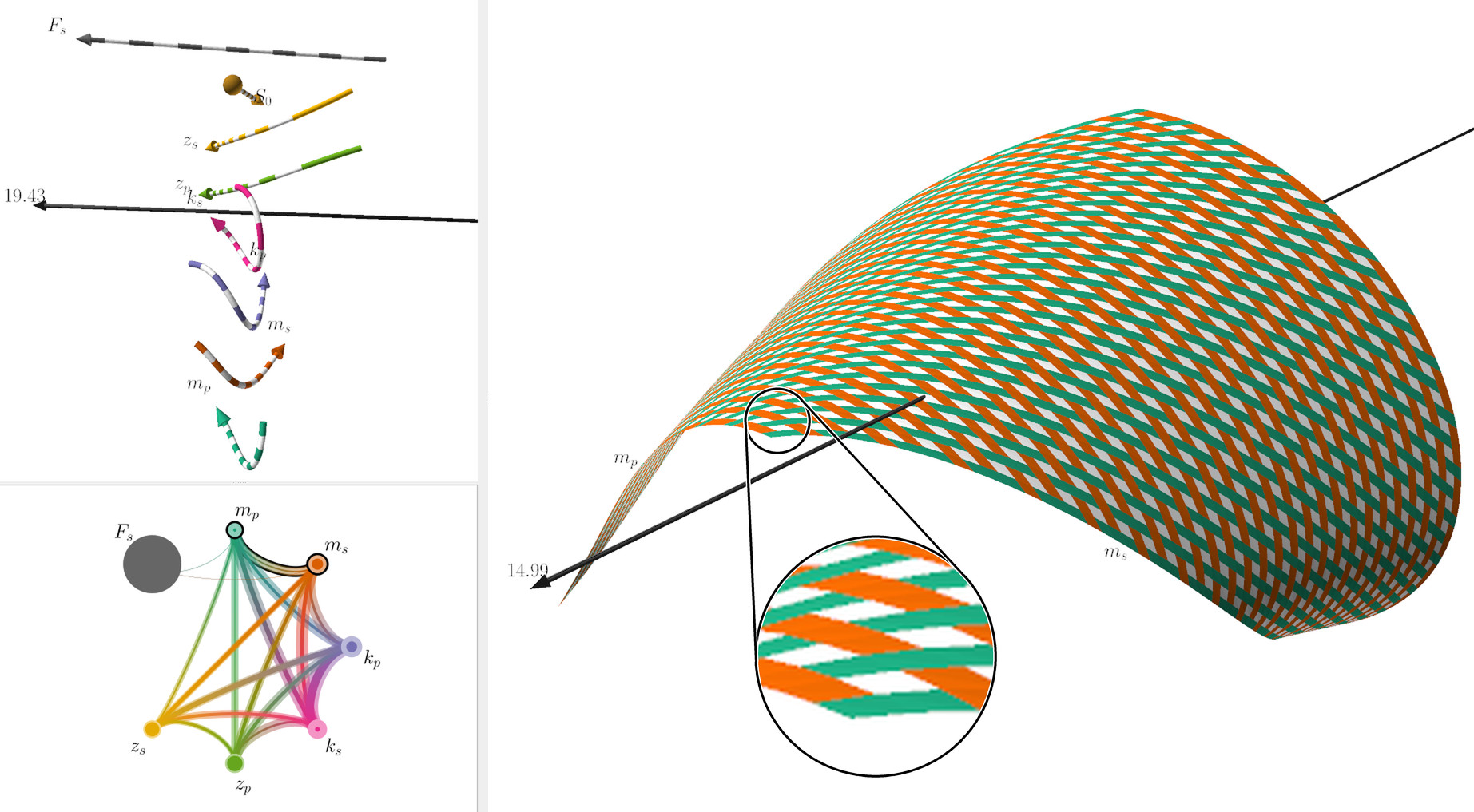}
 \caption{\emph{Damped Oscillator}: we visualize the peak force experienced by a 2-mass oscillating spring system~\cite{DSD:13} depending on the masses $m_p$ and $m_s$. Note the obtuse angles (here shown by the magnifying glass) indicating that those masses mostly cancel each other. The surface's maximum (next to the black arrowhead) is achieved when the primary mass $m_p$ is high and there is little secondary mass $m_s$ to compensate for it.}
 \label{fig:system2}
\end{figure}

\begin{figure}[h] \center
\begin{subfigure}[c]{0.45\columnwidth}
\includegraphics[width=\textwidth]{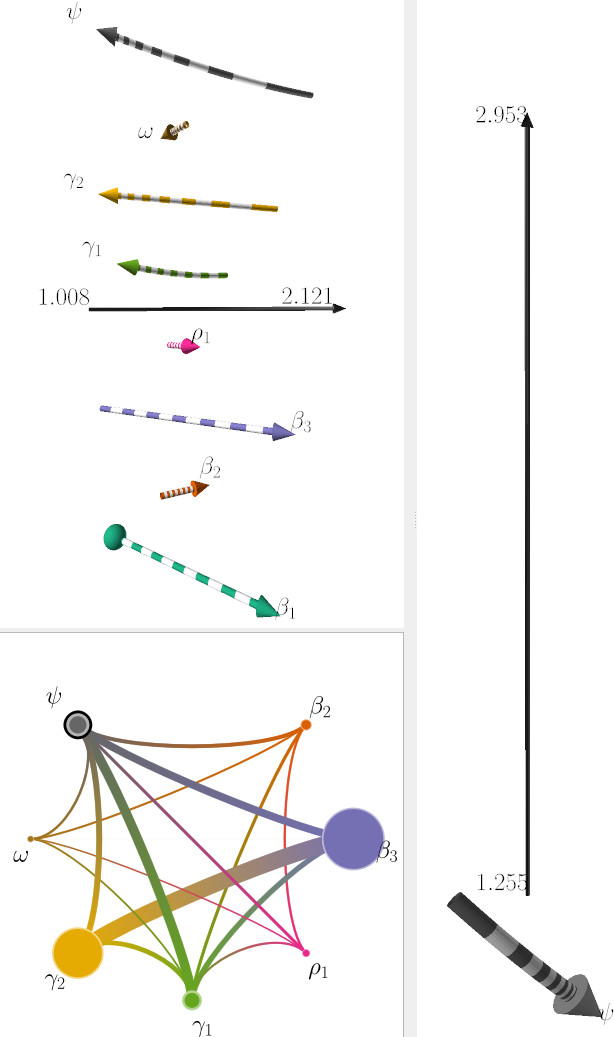}
\caption{Lowest transmission rate $\beta_1$}
\end{subfigure} \hfil
\begin{subfigure}[c]{0.53797\columnwidth}
\includegraphics[width=\textwidth]{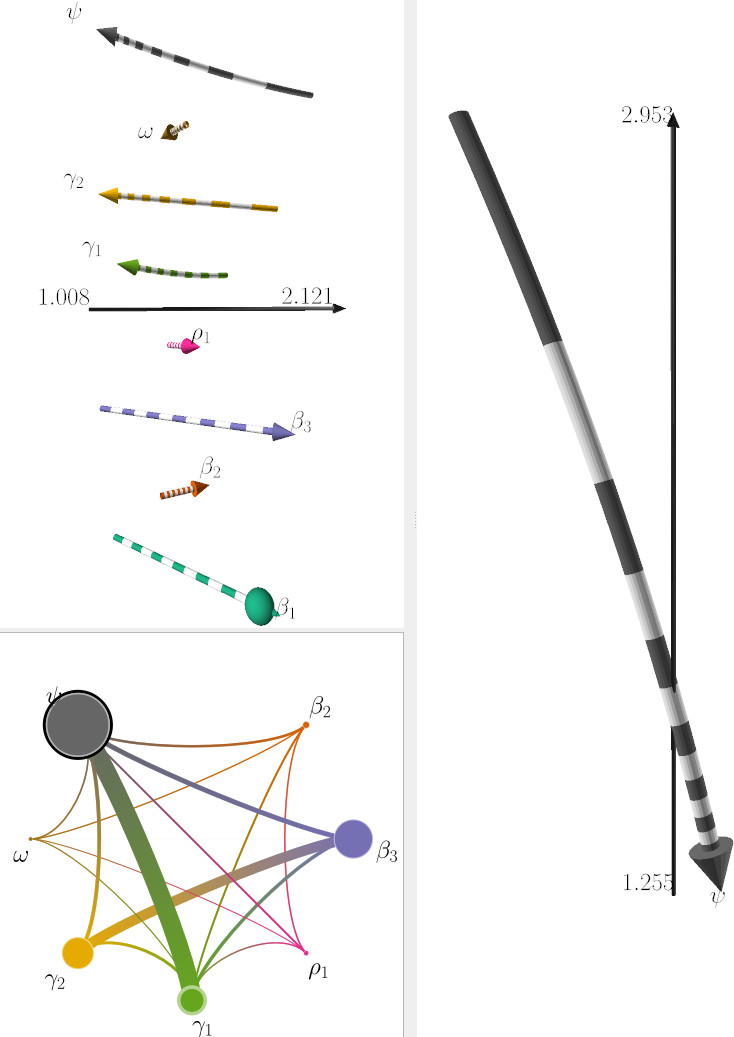}
\caption{Highest transmission rate $\beta_1$}
\end{subfigure}
 \caption{\emph{Ebola Spread}: we vary transmission rate $\beta_1$ (cyan curve; position indicated by the spherical marker) and study its impact on the effectiveness of hospitalization rate $\psi$ (large gray curves) at reducing Ebola infection rate $R$ (black axes). Note that $\psi$ is extremely important at high values of $\beta_1$ (b), where it can halve the overall infection rate. Correspondingly, the gray node in the radial widget in (b) becomes the largest one.}
 \label{fig:system}
\end{figure}

We have combined the features previously discussed into an integrated widget-based visualization application (see Figs.~\ref{fig:system2} and~\ref{fig:system} for example snapshots). Global, single-variable structure is shown via a curve array (Sec.~\ref{sec:curve_array}) within a 3D widget (top left). There is also a 2D minimap widget (bottom left) that starts showing Sobol indices for the overall model, and a 3D widget on the right to show individual parameterized curves and surfaces in detail.

Navigation is governed by an \emph{active tuple} of variables that is empty at the beginning. By clicking a node or an arc on the Sobol minimap, the user can select a variable or pair of variables for further analysis. For instance, if we select an arc connecting an additive and a high-order variable, we expect their joint surface to be mostly rectangularly tiled. The surface will be more or less bent depending on whether there are further high-order interactions with even more variables, as is signaled by the lighter part of their arc.

Then, the corresponding curve or surface visualization is launched in the right 3D widget. The user can also click on the curve array to set a further $n$-th variable to a specific value $x_n := \alpha$, and update this value interactively by sliding a sphere back and forth. Moving the sphere also updates the minimap, which then shows the Sobol indices of the selected 1-variable partial function, as well as whatever curve or surface that is visible in the right widget. This system goes one step further beyond the plot matrix (Sec.~\ref{sec:plot_matrix}) as it can show up to third-order interactions smoothly.

As a last example we use the proposed widget-based tool to explore the \emph{Ebola Spread} model. The most important goal in this problem is ascertaining how resources can be best allocated to reduce the infection rate $R$. With this in mind, the easiest variables to alter in practice are the hospitalization rate $\psi$ and the proper burial rate $\omega$~\cite{DCKJP:18}. Fig.~\ref{fig:system} shows two snapshots of the tool for this model. Note that there are four variables that generally increase the rate $R$ and four that reduce it. An insight that stands out immediately from the curve array in Fig.~\ref{fig:system} is that $\psi$ has a much stronger effect than $\omega$ at reducing the rate $R$. This is precisely one of the main conclusions reached in the study by Diaz et al.~\cite{DCKJP:18}. Furthermore, we can use the proposed widgets to understand under what circumstances this disparity is more or less acute. The model is also interesting for its accelerations and decelerations. For example, variations in the hospitalization rate $\psi$ make much more of a difference for low $\psi$, whereas increasing an already high $\psi$ yields a vanishing improvement only. In addition, the influence of other variables changes drastically when $\psi$ varies. We can examine this scenario in depth by setting $\psi$ to a high value in our application's curve array widget to find out what other parameters would then become useful at further reducing the infection rate.

%------------------------------------------------------------------------------------------------------------------------------
\section{Conclusions}
%------------------------------------------------------------------------------------------------------------------------------

We have contributed a principal component-based dimensionality reduction for global-to-local visualization and sensitivity analysis of dense parameter spaces. To this end we consider the set of all possible partial functions of a model $f$ and project them onto two orthogonal components in the spirit of Sobol's decomposition method for high-dimensional ANOVA. We summarize those components using a few spatial coordinates to form various parameterized manifolds including curves, surfaces, and ensembles thereof. 

In its simplest form, our algorithm boils down to higher-order tensor PCA, on top of which we contributed three conceptual and computational developments:

\begin{itemize}
	\item The abstraction of taking arbitrary partial functions as the set of vectors to project, including those that are defined with respect to groups of variables and thus give rise to multidimensional manifolds;
	\item We split the original $L^2$ space into additive and high-order subspaces so as to separately capture the different kinds of influences that variables (or groups thereof) can have. This gives the proposed mapping a direct interpretation in terms of the ANOVA decomposition and the Sobol indices.
	\item We are aware that computing the principal components of large collections of high-dimensional discretized vectors (with e.g. billions of entries) is a challenging task. We exploited a numerical framework, the tensor train decomposition, that is key to ensure responsiveness and interactivity within the proposed visualization system. The parameter space is cast as a tensor grid and approximated as a low-rank expansion; we extract its principal subspaces directly from the compressed domain.
\end{itemize}

The visualization diagrams made possible by those ingredients are able to readily communicate interactions between up to three input variables. They also provide the user with discriminative information that allows him or her to select interesting combinations of variables as well as specific values for those variables.
We identified how several interesting high-dimensional global and local properties are mapped to specific patterns in 3D curves and shapes, and how individual versus joint-variable effects stand out from our visualization. To the best of our knowledge, this is the first visualization system that is able to communicate such structure in a global-to-local, time-effective manner. 

%------------------------------------------------------------------------------------------------------------------------------
\subsection*{Future Work}

As outlined in the introduction, in this paper we have focused on dense parameter spaces. In particular, no scattered data set (i.e. given collection of samples at fixed locations) was considered as a ground-truth input. Although the domain of application we pursued is attractive, we believe the discrete case remains an equally important target. We believe the proposed method is adaptable to this end: instead of abstracting partial functions over the entire domain, we can show parameterizations that summarize regions or neighborhoods only, for example around feature points or samples from given scattered data. This way we can combine the strengths of global/contextual information (as is only made possible via surrogate modeling) with local structural information as arising from possibly complex sample distributions. We would also like to let users define and move anchor points similarly to interactive PCA~\cite{JZFRC:09}, in order to provide extra layers of informed interaction.

%\acknowledgments{
%The authors wish to thank A, B, and C. This work was supported in part by
%a grant from XYZ (\# 12345-67890).}

%\bibliographystyle{abbrv}
\bibliographystyle{abbrv-doi}

\bibliography{references}

\end{document}